\pdfoutput=1
\documentclass[twocolumn,english,aps,pra,showpacs]{revtex4}
\usepackage[T1]{fontenc}
\usepackage[latin1]{inputenc}
\usepackage{amsmath}
\usepackage{graphicx}
\usepackage{amssymb}
\usepackage{esint}
\usepackage{epsfig}
\usepackage{epstopdf}

\usepackage{color}

\usepackage[normalem]{ulem}

\makeatletter
\@ifundefined{textcolor}{}
{%
 \definecolor{BLACK}{gray}{0}
 \definecolor{WHITE}{gray}{1}
 \definecolor{RED}{rgb}{1,0,0}
 \definecolor{GREEN}{rgb}{0,1,0}
 \definecolor{BLUE}{rgb}{0,0,1}
 \definecolor{CYAN}{cmyk}{1,0,0,0}
 \definecolor{MAGENTA}{cmyk}{0,1,0,0}
 \definecolor{YELLOW}{cmyk}{0,0,1,0}
 }

\@ifundefined{definecolor}{\@ifundefined{definecolor}{\@ifundefined{definecolor}
 {\usepackage{color}}{}
}{}}{}\makeatother

\makeatother

\usepackage{babel}

\makeatother

\usepackage{babel}

\makeatother

\usepackage{babel}

\makeatother

\usepackage{babel}

\makeatother

\usepackage{babel}

\begin{document}

\title{Half-quantum vortex state in a spin-orbit coupled Bose-Einstein condensate}

\author{B. Ramachandhran$^{1}$, Bogdan Opanchuk$^{2}$, Xia-Ji Liu$^{2}$,
Han Pu$^{1}$, Peter D. Drummond$^{2}$, and Hui Hu$^{2}$}

\affiliation{$^{1}$Department of Physics and Astronomy, and Rice Quantum Institute,
Rice University, Houston, TX 77251, USA \\
 $^{2}$ARC Centres of Excellence for Quantum-Atom Optics and Centre
for Atom Optics and Ultrafast Spectroscopy, Swinburne University of
Technology, Melbourne 3122, Australia }

\date{\today}
\begin{abstract}
We investigate theoretically the condensate state and collective excitations
of a two-component Bose gas in two-dimensional harmonic traps subject
to isotropic Rashba spin-orbit coupling. In the weakly interacting
regime when the inter-species interaction is larger than the intra-species
interaction ($g_{\uparrow\downarrow}>g$), we find that the condensate
ground state has a half-quantum-angular-momentum vortex configuration
with spatial rotational symmetry and skyrmion-type spin texture. Upon
increasing the interatomic interaction beyond a threshold $g_{c}$,
the ground state starts to involve higher-order angular momentum components
and thus breaks the rotational symmetry. In the case of $g_{\uparrow\downarrow}<g$,
the condensate becomes unstable towards the superposition of two degenerate
half-quantum vortex states. Both instabilities (at $g>g_{c}$ and
$g_{\uparrow\downarrow}<g$) can be determined by solving the Bogoliubov
equations for collective density oscillations of the half-quantum
vortex state, and by analyzing the softening of mode frequencies.
We present the phase diagram as functions of the interatomic interactions
and the spin-orbit coupling. In addition, we directly simulate the
time-dependent Gross-Pitaevskii equation to examine the dynamical
properties of the system. Finally, we investigate the stability of
the half-quantum vortex state against both the trap anisotropy and
anisotropy in the spin-orbit coupling term. 
\end{abstract}


\maketitle

\section{Introduction}

Owing to the unprecedented control in interatomic interaction, geometry
and purity, atomic quantum gases have proven to be an ideal many-body
platform for exploring fundamental quantum states, such as Bose-Einstein
condensates (BEC) \cite{BEC}, strongly interacting unitary Fermi
superfluids \cite{UnitaryFG1,UnitaryFG2} and Mott-insulating states
\cite{Mott}. One of the latest achievement concerns the spin-orbit
(SO) coupling in an ultracold spinor Bose gas of $^{87}$Rb atoms
\cite{SpielmanNature2011}, induced by the so-called ``synthetic
non-Abelian gauge fields''. Novel quantum states may be anticipated
in the presence of SO coupling \cite{GalitskiPRA2008,ChiralConfinement,ZhaiPRL2010,HoPreprint,ZhangPreprint,YouPRA2011,MachidaPRA2011,WuCPL2011,HanPreprint,WuPreprint,GalitskiPreprint,HuPreprint,SinhaPreprint}.
Indeed, for a homogeneous SO coupled spin-1/2 Bose gas with intra-
and inter-species interactions ($g$ and $g_{_{\uparrow\downarrow}}$),
a single plane-wave or a density-stripe condensate state has been
predicted \cite{ZhaiPRL2010}, depending on whether $g$ is smaller
or larger than $g_{_{\uparrow\downarrow}}$. Interesting density patterns
have been observed in the theoretical simulations for an SO coupled
spinor condensate, in the absence \cite{ZhaiPRL2010,YouPRA2011,MachidaPRA2011,HuPreprint,SinhaPreprint}
or presence \cite{HanPreprint,WuPreprint,GalitskiPreprint} of rotation.
The phenomenon of self-trapped BECs has also been proposed, in particular,
in one-dimensional (1D) geometry \cite{ChiralConfinement}.

\begin{figure}[htp]
\begin{centering}
\includegraphics[clip,width=0.48\textwidth]{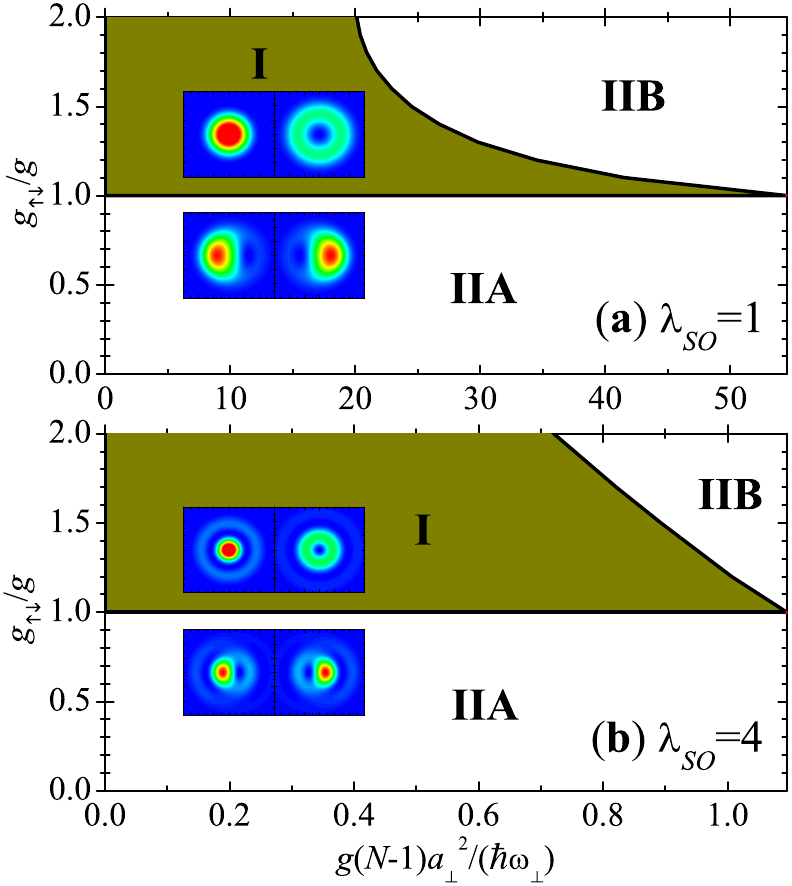} 
\par\end{centering}

\caption{(color online). Phase diagram at two dimensionless SO coupling strengths,
$\lambda_{SO}=1$ (a) and $\lambda_{SO}=4$ (b). The half-quantum
vortex state (the phase I) becomes unstable when the intra-species
interaction is larger than the inter-species interaction ($g>g_{_{\uparrow\downarrow}}$,
the phase IIA) or when the interatomic interactions are sufficient
strong ($g>g_{c}$, the phase IIB). The insets shows the density patterns
of the spin-up and spin-down bosons in the phases I and IIA. We note
that, the critical interaction strength $g_{c}$ increases rapidly
with decreasing the SO coupling strength $\lambda_{SO}$.}

\label{fig1} 
\end{figure}

In this work, we show that in a Rashba SO coupled, weakly interacting
spin-1/2 Bose gas in two-dimensional (2D) harmonic traps,
all bosons may condense into a non-trivial half-integer angular momentum
state (or a half-quantum vortex state) with a skyrmion-type spin texture.
We solve the mean-field Gross-Pitaevskii equation (GPE) for its density
distributions and spin textures, and obtain its collective excitation
spectrum by solving the Bogoliubov equation and by directly simulating
real-time propagation of the GPE ground state under perturbation.
The condensation of an SO coupled spin-1/2 Bose gas into a half-quantum
vortex configuration was first suggested by Congjun Wu and co-workers
in 2008 and its existence was discussed under the condition that the
interaction is SU(2) symmetric, i.e., $g=g_{_{\uparrow\downarrow}}$
\cite{WuCPL2011}. Here, we explore systematically the parameter space
for the half-quantum vortex state and analyze its stability. We present
a phase diagram for the half-quantum vortex state as functions of
the SO coupling and the interatomic interaction strengths. We also
investigate the dynamical properties of the half-quantum vortex state
by directly simulating the time-dependent GPE. Finally, the stability
of the half-quantum vortex state against both the trap anisotropy
and anisotropy in the spin-orbit coupling term is examined.

Our main results are summarized in Fig. \ref{fig1}. The half-quantum
vortex state (the phase I) is the ground state if the intra-species
interaction is smaller than the inter-species interaction ($g<g_{_{\uparrow\downarrow}}$)
and if the interaction strength is below a threshold ($g<g_{_{c}}$).
Otherwise, it becomes energetically unstable towards a superposition
state of two degenerate half-quantum vortex states (the phase IIA),
or a state involving higher-order angular momentum components (the
phase IIB). With decreasing the dimensionless SO coupling strength
$\lambda_{SO}$, the threshold $g_{_{c}}$ becomes exponentially large,
leading to a large parameter space for the half-quantum vortex state
(see Fig. \ref{fig11}). It is therefore feasible to be observed in
the current experiments with ultracold SO coupled spinor Bose gases
of $^{87}$Rb atoms.

The rest of the paper is organized as follows. In the next section,
we outline the model Hamiltonian and discuss briefly the existence
of half-quantum vortex state in the non-interacting limit. In Sec.
III, we present the numerical procedure of solving the GPE and Bogoliubov
equations and discuss the typical density distributions and collective
mode behaviors of the half-quantum vortex state. The collective excitation
spectrum obtained from the Bogoliubov equation is compared to a direct
simulation of the time-dependent GPE. In Sec. IV, we analyze the stability
of the half-quantum vortex state by monitoring the softening of collective
mode frequencies and by comparing the energy with that of some competing
states. The phase diagram is then constructed as functions of interatomic
interactions and SO coupling. The stability against the anisotropy
in trapping potential and in spin-orbit coupling term is also carefully
examined. Finally, we summarize in Sec. V and give some concluding
remarks.

\section{Theoretical framework}

We consider a two-component Bose gas confined in a 2D isotropic harmonic
trap $V(\rho)=M\omega_{\perp}^{2}(x^{2}+y^{2})/2=M\omega_{\perp}^{2}\rho^{2}/2$
with a Rashba SO coupling ${\cal V}_{SO}=-i\lambda_{R}(\hat{\sigma}_{x}\partial_{y}-\hat{\sigma}_{y}\partial_{x})$,
where $\lambda_{R}$ is the Rashba SO coupling strength and $\hat{\sigma}_{x}$,
$\hat{\sigma}_{y}$, and $\hat{\sigma}_{z}$ are the $2\times2$ Pauli
matrices. The model Hamiltonian ${\cal H=}\int d{\bf r[}{\cal H}_{0}+{\cal H}_{{\rm int}}]$
is given by, 
\begin{eqnarray}
{\cal H}_{0} & = & \Psi^{\dagger}\left[-\frac{\hbar^{2}\nabla^{2}}{2M}+V\left(\rho\right)+{\cal V}_{SO}-\mu\right]\Psi{\bf ,}\\
{\cal H}_{{\rm int}} & = & (g/2)\sum_{\sigma=\uparrow,\downarrow}\Psi_{\sigma}^{\dagger}\Psi_{\sigma}^{\dagger}\Psi_{\sigma}\Psi_{\sigma}{\bf +}g_{\uparrow\downarrow}\Psi_{\uparrow}^{\dagger}\Psi_{\uparrow}\Psi_{\downarrow}^{\dagger}\Psi_{\downarrow}{\bf ,}
\end{eqnarray}
 where ${\bf r}=(x,y)$ and $\Psi=[\Psi_{\uparrow}({\bf r)},\Psi_{\downarrow}({\bf r)}]^{T}$
denotes the spinor Bose field operators in a collective way, and the
chemical potential $\mu$ is to be determined by the total number
of bosons $N$, i.e., $\int d{\bf r}\Psi^{\dagger}\Psi=N$. For simplicity,
we have assumed equal intra-species interaction strength $g_{\uparrow\uparrow}=g_{\downarrow\downarrow}=g$.
In experiments, the two-dimensionality can be readily realized by
imposing a strong harmonic potential $V(z)=M\omega_{z}^{2}z^{2}/2$
along axial direction, in such a way that $\mu,k_{B}T\ll\hbar\omega_{z}$
\cite{Dalibard2D}. For the realistic case of $^{87}$Rb atoms, the
interaction strengths can be calculated from the two \textit{s}-wave
scattering lengths $a\simeq100a_{B}$ and $a_{\uparrow\downarrow}$,
using $g=\sqrt{8\pi}(\hbar^{2}/M)(a/a_{z})$ and $g_{\uparrow\downarrow}=\sqrt{8\pi}(\hbar^{2}/M)(a_{\uparrow\downarrow}/a_{z})$,
respectively. Here $a_{z}=\sqrt{\hbar/(M\omega_{z})}$ is the characteristic
oscillator length in $z$-direction.

For a weakly interacting Bose gas at zero temperature, we assume that
all the bosons condense into a single quantum state $\Phi({\bf r)=}[\Phi_{\uparrow}({\bf r)},\Phi_{\downarrow}({\bf r)}]^{T}$.
Following the standard mean-field theory \cite{AllanPRB1996}, we
separate the field operator into a condensate and a fluctuation part,
$\Psi_{\sigma}({\bf r)=}\Phi_{\sigma}({\bf r)+}\tilde{\Psi}_{\sigma}({\bf r)}$.
Keeping up to the quadratic terms in $\tilde{\Psi}_{\sigma}({\bf r)}$,
this separation leads to ${\cal H=}\int d{\bf r[}{\cal H}_{{\rm GP}}+{\cal H}_{T}]$,
where the condensate part is given by, 
\begin{eqnarray}
{\cal H}_{{\rm GP}} & = & \Phi^{\dagger}\left[{\cal H}_{{\rm osc}}+{\cal V}_{SO}-\mu\right]\Phi\nonumber \\
 &  & +\frac{g}{2}\left(\left|\Phi_{\uparrow}\right|^{4}+\left|\Phi_{\downarrow}\right|^{4}\right)+g_{\uparrow\downarrow}\left|\Phi_{\uparrow}\Phi_{\downarrow}\right|^{2},\label{HGP}
\end{eqnarray}
 and the fluctuation part ${\cal H}_{T}=\tilde{\Psi}^{\dagger}{\cal H}_{{\rm Bog}}\tilde{\Psi}$
with \begin{widetext} 
\begin{equation}
{\cal H}_{{\rm Bog}}=\left[\begin{array}{cccc}
{\cal H}_{s_{\uparrow}}+g\left|\Phi_{\uparrow}\right|^{2} & V_{{\rm so}}+g_{\uparrow\downarrow}\Phi_{\uparrow}\Phi_{\downarrow}^{*} & g\Phi_{\uparrow}^{2} & g_{\uparrow\downarrow}\Phi_{\uparrow}\Phi_{\downarrow}\\
V_{{\rm so}}^{\dagger}+g_{\uparrow\downarrow}\Phi_{\uparrow}^{*}\Phi_{\downarrow} & {\cal H}_{s_{\downarrow}}+g\left|\Phi_{\downarrow}\right|^{2} & g_{\uparrow\downarrow}\Phi_{\uparrow}\Phi_{\downarrow} & g\Phi_{\downarrow}^{2}\\
g\left(\Phi_{\uparrow}^{*}\right)^{2} & g_{\uparrow\downarrow}\Phi_{\uparrow}^{*}\Phi_{\downarrow}^{*} & {\cal H}_{s_{\uparrow}}+g\left|\Phi_{\uparrow}\right|^{2} & -V_{{\rm so}}^{\dagger}+g_{\uparrow\downarrow}\Phi_{\uparrow}^{*}\Phi_{\downarrow}\\
g_{\uparrow\downarrow}\Phi_{\uparrow}^{*}\Phi_{\downarrow}^{*} & g\left(\Phi_{\downarrow}^{*}\right)^{2} & -V_{{\rm so}}+g_{\uparrow\downarrow}\Phi_{\uparrow}\Phi_{\downarrow}^{*} & {\cal H}_{s_{\downarrow}}+g\left|\Phi_{\downarrow}\right|^{2}
\end{array}\right].
\end{equation}
 \end{widetext} Here ${\cal H}_{{\rm osc}}\equiv-\hbar^{2}\nabla^{2}/(2M)+V\left(\rho\right)$,
${\cal H}_{s_{\uparrow}}\equiv{\cal H}_{{\rm osc}}+g\left|\Phi_{\uparrow}\right|^{2}+g_{\uparrow\downarrow}\left|\Phi_{\downarrow}\right|^{2}-\mu$
and ${\cal H}_{s_{\downarrow}}\equiv{\cal H}_{{\rm osc}}+g_{\uparrow\downarrow}\left|\Phi_{\uparrow}\right|^{2}+g\left|\Phi_{\downarrow}\right|^{2}-\mu$,
$V_{{\rm so}}\equiv-i\lambda_{R}(\partial_{y}+i\partial_{x})$ and
$V_{{\rm so}}^{\dagger}\equiv-i\lambda_{R}(\partial_{y}-i\partial_{x})$,
and we have introduced a $4\times4$ Nambu spinor $\tilde{\Psi}=[\tilde{\Psi}_{\uparrow}({\bf r)},\tilde{\Psi}_{\downarrow}({\bf r),}\tilde{\Psi}_{\uparrow}^{\dagger}({\bf r)},\tilde{\Psi}_{\downarrow}^{\dagger}({\bf r)}]^{T}$.

The condensate wave-function can be obtained from the GP equations
$\delta{\cal H}_{{\rm GP}}/\delta\Phi({\bf r})=0$ \cite{AllanPRB1996},
or explicitly, 
\begin{equation}
\left[\begin{array}{cc}
{\cal H}_{s_{\uparrow}} & -i\lambda_{R}(\partial_{y}+i\partial_{x})\\
-i\lambda_{R}(\partial_{y}-i\partial_{x}) & {\cal H}_{s_{\downarrow}}
\end{array}\right]\left[\begin{array}{c}
\Phi_{\uparrow}\left({\bf r}\right)\\
\Phi_{\downarrow}\left({\bf r}\right)
\end{array}\right]=0\text{.}\label{GP}
\end{equation}
 At zero temperature, we assume a single condensate state with {\em
zero} quantum depletion, so that the condensate wave-function is
normalized by $\int d{\bf r}[\left|\Phi_{\uparrow}\right|^{2}+\left|\Phi_{\downarrow}\right|^{2}]=N$,
where $N$ is the total number of bosons. The equation becomes simplified
if we write $\Phi_{\uparrow}=N^{1/2}\phi_{\uparrow}$ and $\Phi_{\downarrow}=N^{1/2}\phi_{\downarrow}$
and use accordingly the interaction strengths $g(N-1)$ and $g_{\uparrow\downarrow}(N-1)$.
The normalization condition becomes $\int d{\bf r}\,[\left|\phi_{\uparrow}\right|^{2}+\left|\phi_{\downarrow}\right|^{2}]=1$.

The quasi-particle wave-functions with energy $\hbar\omega$ satisfy
the Bogoliubov equations \cite{AllanPRB1996}, 
\begin{equation}
{\cal H}_{Bog}\left[\begin{array}{c}
u_{\uparrow}\left({\bf r}\right)\\
u_{\downarrow}\left({\bf r}\right)\\
v_{\uparrow}\left({\bf r}\right)\\
v_{\downarrow}\left({\bf r}\right)
\end{array}\right]=\hbar\omega\left[\begin{array}{c}
+u_{\uparrow}\left({\bf r}\right)\\
+u_{\downarrow}\left({\bf r}\right)\\
-v_{\uparrow}\left({\bf r}\right)\\
-v_{\downarrow}\left({\bf r}\right)
\end{array}\right],\label{Bog}
\end{equation}
 and is normalized by $\int d{\bf r}[\left|u_{\uparrow}\right|^{2}+\left|u_{\downarrow}\right|^{2}-\left|v_{\uparrow}\right|^{2}-\left|v_{\downarrow}\right|^{2}]=1$.
These Bogoliubov quasi-particles correspond to the different collective
density oscillation modes around the condensate with the frequency
$\omega$~\cite{LiuPRA2004}. It is easy to see that the wave-function
$[v_{\uparrow}^{*}\left({\bf r}\right),v_{\downarrow}^{*}\left({\bf r}\right),u_{\uparrow}^{*}\left({\bf r}\right),u_{\downarrow}^{*}\left({\bf r}\right)]^{T}$
is also a solution of Eq.~(\ref{Bog}), but with energy $-\hbar\omega$.
This is anticipated for the usual Bogoliubov transformation. Physically,
we should restrict to a non-negative mode frequency, $\omega\geq0$.

In harmonic traps, it is natural to use the trap units, i.e. to take
$\hbar\omega_{\perp}$ as the unit for energy and the harmonic oscillator
length $a_{\perp}=\sqrt{\hbar/(M\omega_{\perp})}$ as the unit for
length. This is equivalent to set $\hbar=k_{B}=M=\omega_{\perp}=1$.
For the SO coupling, we introduce an SO coupling length $a_{\lambda}=\hbar^{2}/(M\lambda_{R})$
and consequently define a dimensionless SO coupling strength $\lambda_{SO}=a_{\perp}/a_{\lambda}=\sqrt{(M/\hbar^{3})}\lambda_{R}/\sqrt{\omega_{\perp}}$.
In an SO coupled spin-1/2 BEC of $^{87}$Rb atoms as realized recently
by the NIST group \cite{SpielmanNature2011}, $\lambda_{SO}$ is about
$10$. In the typical experiment for 2D spin-1/2 $^{87}$Rb BECs \cite{Dalibard2D},
the interatomic interaction strengths are about $g(N-1)\approx g_{\uparrow\downarrow}(N-1)=10^{2}\sim10^{3}\hbar\omega_{\perp}/a_{\perp}^{2}$.
These coupling strengths, however, can be precisely tuned by properly
choosing the parameters of the laser fields that lead to the harmonic
confinement and the SO coupling.

\subsection{Single-particle solutions}

The appearance of the half-quantum vortex state may be easily understood
in the non-interacting limit \cite{WuCPL2011}. In the absence of
interatomic interactions, the single-particle wave-function $[\phi_{\uparrow}\left({\bf r}\right),\phi_{\downarrow}\left({\bf r}\right)]^{T}$
with energy $\epsilon$ is given by, 
\begin{equation}
\left[\begin{array}{cc}
{\cal H}_{osc} & -i\lambda_{R}(\partial_{y}+i\partial_{x})\\
-i\lambda_{R}(\partial_{y}-i\partial_{x}) & {\cal H}_{osc}
\end{array}\right]\left[\begin{array}{c}
\phi_{\uparrow}\\
\phi_{\downarrow}
\end{array}\right]=\epsilon\left[\begin{array}{c}
\phi_{\uparrow}\\
\phi_{\downarrow}
\end{array}\right]\text{.}
\end{equation}
 In polar coordinates ($\rho,\varphi$), we have $-i(\partial_{y}\pm i\partial_{x})=e^{\mp i\varphi}[\pm\partial/\partial\rho-(i/\rho)\partial/\partial\varphi]$.
Because of the isotropic harmonic potential $V\left(\rho\right)$,
the single-particle wave-function may have a well-defined azimuthal
angular momentum $l_{z}=m$ and may take the form, 
\begin{equation}
\phi_{m}({\bf r})=\left[\begin{array}{c}
\phi_{\uparrow}(\rho)\\
\phi_{\downarrow}(\rho)e^{i\varphi}
\end{array}\right]\frac{e^{im\varphi}}{\sqrt{2\pi}}.
\end{equation}
 This state also has a well-defined total angular momentum $j_{z}=l_{z}+s_{z}=m+1/2$.
In general, we may denote the energy spectrum as $\epsilon_{nm}$,
where $n=(0,1,2...)$ is the quantum number for the transverse (radial)
direction. There is an interesting two-fold degeneracy of the energy
spectrum: any eigenstate $\phi({\bf r})=[\phi_{\uparrow}({\bf r}),\phi_{\downarrow}({\bf r})]^{T}$
is degenerate with its time-reversal partner ${\cal T}\phi({\bf r})\equiv(i\sigma_{y}{\cal C})\phi({\bf r})=$
$[\phi_{\downarrow}^{*}({\bf r}),-\phi_{\uparrow}^{*}({\bf r})]^{T}$.
Here ${\cal C}$ is the complex conjugate operation. This Kramer doublet
is the direct consequence of the time-reversal symmetry satisfied
by the model Hamiltonian. It preserves as well in the presence of
interatomic interactions. As a result, we may restrict the quantum
number $m$ to be non-negative integers, as a negative $m$ can always
be regarded as the time-reversal partner for a state with $m\geq0$.

To solve numerically the single-particle spectrum, we adopt a basis-expansion
method. To this end, we expand first, 
\begin{eqnarray}
\phi_{\uparrow}(\rho) & = & \sum_{k}A_{k}R_{km}\left(\rho\right),\\
\phi_{\downarrow}(\rho) & = & \sum_{k}B_{k}R_{km+1}\left(\rho\right),
\end{eqnarray}
 where 
\begin{equation}
R_{km}=\frac{1}{a_{\perp}}\sqrt{\frac{2k!}{\left(k+\left|m\right|\right)!}}\left(\frac{\rho}{a_{\perp}}\right)^{\left|m\right|}e^{-\frac{\rho^{2}}{2a_{\perp}^{2}}}{\cal L}_{k}^{\left|m\right|}(\frac{\rho^{2}}{a_{\perp}^{2}})
\end{equation}
 is the radial wave-function of a 2D harmonic oscillator ${\cal H}_{osc}$
with energy $(2k+\left|m\right|+1)\hbar\omega_{\perp}$, and ${\cal L}_{k}^{\left|m\right|}$
is the associated Legendre polynomial. Then, we have the following
secular matrix, 
\begin{equation}
\left[\begin{array}{cc}
{\cal H}_{{\rm osc\uparrow}} & {\cal M}^{T}\\
{\cal M} & {\cal H}_{{\rm osc\downarrow}}
\end{array}\right]\left[\begin{array}{c}
A_{k}\\
B_{k}
\end{array}\right]=\epsilon\left[\begin{array}{c}
A_{k}\\
B_{k}
\end{array}\right],\label{spMatrix}
\end{equation}
 where the matrix elements are given by (for $m\geq0$) 
\begin{eqnarray*}
{\cal H}_{{\rm osc\uparrow,kk^{\prime}}} & = & \hbar\omega_{\perp}\left[2k+m+1\right]\delta_{kk^{\prime}},\\
{\cal H}_{{\rm osc\downarrow,kk^{\prime}}} & = & \hbar\omega_{\perp}\left[2k+\left(m+1\right)+1\right]\delta_{kk^{\prime}},\\
{\cal M}_{kk^{\prime}} & = & \hbar\omega_{\perp}{\lambda_{SO}}\left[\sqrt{k^{\prime}+m+1}\delta_{kk^{\prime}}+\sqrt{k^{\prime}}\delta_{kk^{\prime}-1}\right].
\end{eqnarray*}
 Diagonalization of the secular matrix Eq. (\ref{spMatrix}) leads
to the single-particle spectrum and single-particle wave-functions.
In numerical calculations, it is necessary to impose a cut-off $k_{\max}$
for the radial quantum number $k$ of the 2D harmonic oscillator.
For $\lambda_{SO}\leq20$, we find that $k_{\max}=256$ is already
sufficiently large to have an accurate energy spectrum. With this
cut-off, the dimension of the secular matrix in Eq. (\ref{spMatrix})
is $2k_{\max}=512$.

\begin{figure}[htp]
\begin{centering}
\includegraphics[clip,width=0.48\textwidth]{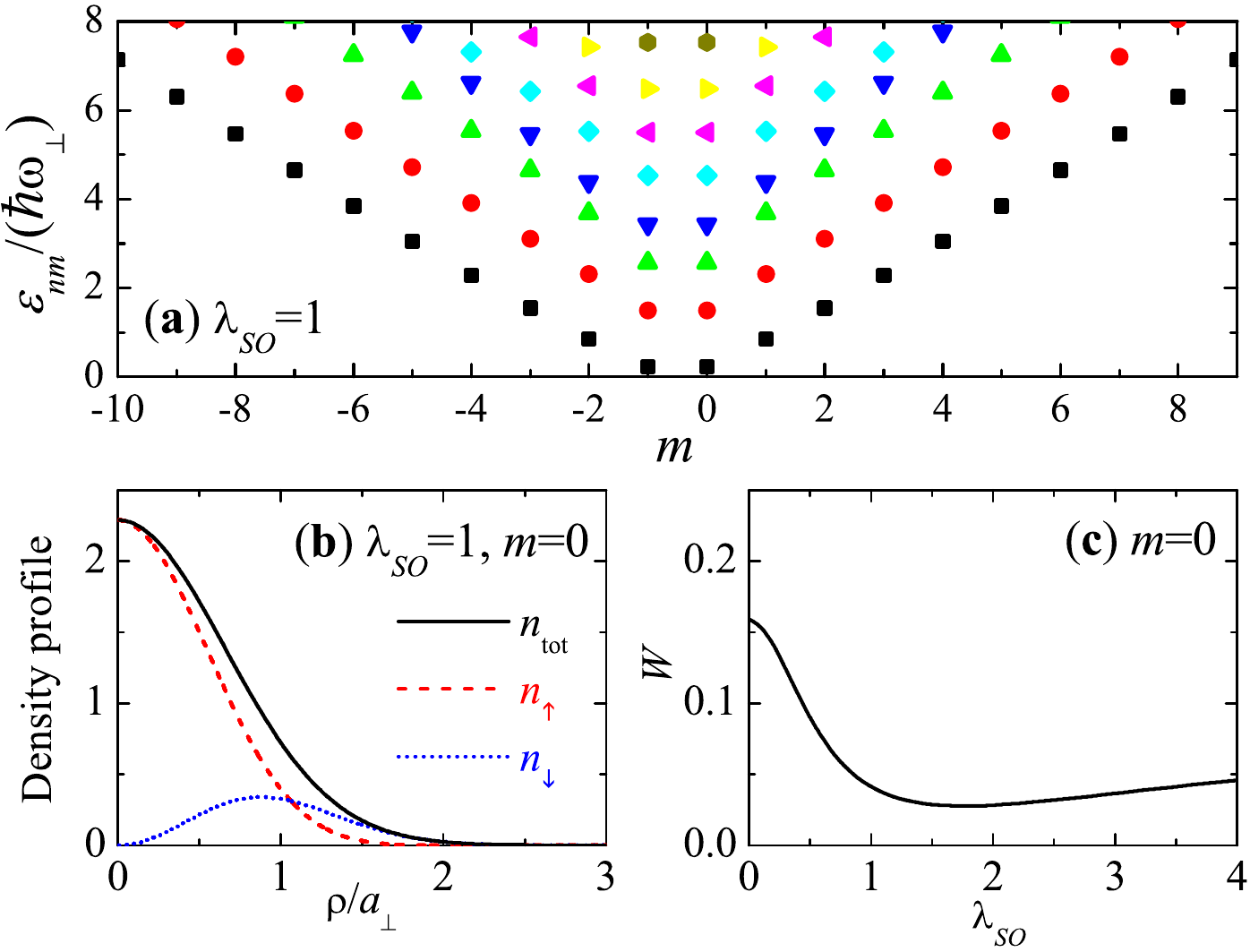} 
\par\end{centering}

\caption{(color online). (a) Single-particle energy spectrum at $\lambda_{SO}=1$.
(b) The density profiles for the single-particle state with $m=0$
at $\lambda_{SO}=1$. (c) The $W$-function for the $m=0$ single-particle
state as a function of SO coupling strength. It is always positive
at arbitrary SO coupling strength.}

\label{fig2} 
\end{figure}

In Fig. 2a, we show the single-particle energy spectrum at $\lambda_{SO}=1$.
For arbitrary SO\ interaction strength, we find numerically that
the doublet single-particle ground state always occurs at $m=0$ (or
$m=-1$ for its time-reversal partner state).

\subsection{Appearance of the half-quantum vortex state}

The single-particle state with $m=0$, $\phi_{0}({\bf r})=[\phi_{\uparrow}(\rho),\phi_{\downarrow}(\rho)e^{i\varphi}]^{T}/\sqrt{2\pi}$,
has a half-quantum vortex configuration \cite{WuCPL2011,HQVS}, as
the spin-up component stays in the $s$-state while the spin-down
component in the $p$-state and the resulting spin texture is of skyrmion
type (see Fig. 2b for density distributions and Sec. IIIB for more
discussions on spin-texture). In the absence of interactions, however,
there is a degenerate time-reversal state, ${\cal T}\phi_{0}({\bf r})=[\phi_{\downarrow}(\rho)e^{-i\varphi},-\phi_{\uparrow}(\rho)]^{T}/\sqrt{2\pi}$,
which is also a half-quantum vortex state. Therefore, in general,
the ground single-particle state is a superposition of two degenerate
half-quantum vortex states of $\phi_{0}({\bf r})$ and ${\cal T}\phi_{0}({\bf r})$,
which takes the form $\phi_{s}({\bf r})=\alpha\phi_{0}({\bf r})+\beta{\cal T}\phi_{0}({\bf r})$,
or explicitly, 
\begin{equation}
\phi_{s}({\bf r})=\frac{1}{\sqrt{2\pi}}\left[\begin{array}{c}
\alpha\phi_{\uparrow}(\rho)+\beta\phi_{\downarrow}(\rho)e^{-i\varphi}\\
\alpha\phi_{\downarrow}(\rho)e^{i\varphi}-\beta\phi_{\uparrow}(\rho)
\end{array}\right].
\end{equation}
 Here $\alpha$ and $\beta$ are two arbitrary complex numbers satisfying
$\left|\alpha\right|^{2}+\left|\beta\right|^{2}=1$.

In the presence of very weak interatomic interactions such that $g(N-1)a_{\perp}^{2},g_{\uparrow\downarrow}(N-1)a_{\perp}^{2}\ll\Delta\epsilon$,
where $\Delta\epsilon$ is the energy difference between the single-particle
ground state $\phi_{0}({\bf r})$ and the first excited state $\phi_{1}({\bf r})$,
we may determine the superposition coefficients $\alpha$ and $\beta$
by minimizing the GP energy, $E_{{\rm GP}}[\phi_{s}({\bf r})]=\int d{\bf r}\,{\cal H}_{{\rm GP}}[\phi_{s}({\bf r})]$.
After a simple algebra, we find that, 
\begin{eqnarray}
\Delta E & = & E_{{\rm GP}}[\phi_{s}({\bf r})]-E_{{\rm GP}}[\phi_{0}({\bf r})],\\
 & = & \left(g_{\uparrow\downarrow}-g\right)(N-1)\left|\alpha\beta\right|^{2}W[\phi_{0}({\bf r})],
\end{eqnarray}
 where the $W$-function is given by, 
\begin{equation}
W[\phi({\bf r})]=\int d{\bf r}[(\left|\phi_{\uparrow}\right|^{2}-\left|\phi_{\downarrow}\right|^{2})^{2}-2\phi_{\uparrow}^{2}\phi_{\downarrow}^{2}].
\end{equation}
 Therefore, a half-quantum vortex state is preferable if $(g_{\uparrow\downarrow}-g)W>0$.
Otherwise, an equal-weight superposition of two degenerate half-quantum
vortex states with $\left|\alpha\right|=\left|\beta\right|=1/\sqrt{2}$
will be the ground state. As shown in Fig. 2c, the $W$-function for
$\phi_{0}({\bf r})$ is positive for {\em arbitrary} SO coupling.
We thus conclude that a half-quantum vortex state should appear at
weak interatomic interactions provided that the inter-species interaction
is larger than the intra-species interaction ($g_{\uparrow\downarrow}>g$).

\section{Density distributions and collective excitations}

Let us now consider finite interatomic interactions, by solving the
GPE for density distributions and spin-textures, and the Bogoliubov
equation for the collective density excitations.

\subsection{GPE solutions of the half-quantum vortex state}

For the half-quantum vortex condensate state with $m=0$, the GP equation
becomes ${\cal L}_{GP}[\phi_{\uparrow}\left(\rho\right),\phi_{\downarrow}\left(\rho\right)]=0$,
where 
\begin{equation}
{\cal L}_{GP}=\left[\begin{array}{cc}
{\cal H}_{s,0}+\bar{g}\phi_{\uparrow}^{2}+\bar{g}_{\uparrow\downarrow}\phi_{\downarrow}^{2} & \lambda_{R}\left(\partial_{\rho}+1/\rho\right)\\
\lambda_{R}\left(-\partial_{\rho}\right) & {\cal H}_{s,1}+\bar{g}_{\uparrow\downarrow}\phi_{\uparrow}^{2}+\bar{g}\phi_{\downarrow}^{2}
\end{array}\right],
\end{equation}
 $\bar{g}\equiv g(N-1)/(2\pi)$ and $\bar{g}_{\uparrow\downarrow}\equiv g_{\uparrow\downarrow}(N-1)/(2\pi)$,
and ${\cal H}_{s,m}\equiv-[\hbar^{2}/(2M)][\partial^{2}/\partial\rho^{2}+(1/\rho)\partial_{\rho}-m^{2}/\rho^{2}]+V(\rho)-\mu$.
The numerical procedure for solving GPE is very similar to that for
single-particle states in Eq. (\ref{spMatrix}). We expand $\phi_{\uparrow}(\rho)=\sum_{k}A_{k}R_{k0}\left(\rho\right)$
and $\phi_{\downarrow}(\rho)=\sum_{k}B_{k}R_{k1}\left(\rho\right)$,
and obtain the secular matrix (with $m=0$), 
\begin{equation}
\left[\begin{array}{cc}
{\cal H}_{osc\uparrow}+{\cal I}_{\uparrow} & {\cal M}^{T}\\
{\cal M} & {\cal H}_{osc\downarrow}+{\cal I}_{\downarrow}
\end{array}\right]\left[\begin{array}{c}
A_{k}\\
B_{k}
\end{array}\right]=\mu\left[\begin{array}{c}
A_{k}\\
B_{k}
\end{array}\right],
\end{equation}
 where 
\begin{eqnarray}
{\cal I}_{\uparrow,kk^{\prime}} & = & \int_{0}^{\infty}\rho d\rho R_{k0}\left(\rho\right)\left(\bar{g}\phi_{\uparrow}^{2}+\bar{g}_{\uparrow\downarrow}\phi_{\downarrow}^{2}\right)R_{k^{\prime}0}\left(\rho\right),\\
{\cal I}_{\downarrow,kk^{\prime}} & = & \int_{0}^{\infty}\rho d\rho R_{k1}\left(\rho\right)\left(\bar{g}_{\uparrow\downarrow}\phi_{\uparrow}^{2}+\bar{g}\phi_{\downarrow}^{2}\right)R_{k^{\prime}1}\left(\rho\right).
\end{eqnarray}
 The chemical potential is given by the lowest eigenvalue of the secular
matrix. Due to the non-linear terms of ${\cal I}_{\uparrow,kk^{\prime}}$
and ${\cal I}_{\uparrow,kk^{\prime}}$, we have to update the condensate
wave-functions and densities iteratively. To overcome the large non-linearity,
we use a simple mixing scheme by setting a small parameter $0<\gamma<1$
and replace the previous density $\phi_{\sigma,old}^{2}$ by $\left(1-\gamma\right)\phi_{\sigma,old}^{2}+\gamma\phi_{\sigma}^{2}$,
where $\phi_{\sigma}^{2}$ is the density calculated in the current
step \cite{PuPRL1998}. The choice of $\gamma$ depends on the interaction
strengths. It becomes smaller for larger $\bar{g}$ and $\bar{g}_{\uparrow\downarrow}$.
We run the iteration until convergence is achieved within a set tolerance.
We have checked that this procedure of solving GPE is stable for interaction
strengths up to $g(N-1),g_{\uparrow\downarrow}(N-1)<10^{3}\hbar\omega_{\perp}/a_{\perp}^{2}$.
For even larger non-linearity, it seems to be impractical to expand
the condensate wave-function using the 2D harmonic oscillator basis.
Therefore for large interaction strengths, we use a time-splitting
spectral method (TSSP) technique to solve the coupled GP equations
and obtain the ground state by imaginary-time propagation \cite{Bao200304,WangJCAM2006}.
For small interaction strengths, results obtained from TSSP are identical
to those obtained from the basis-expansion method.

\subsection{Density distributions and spin textures}

\begin{figure}[htp]
\begin{centering}
\includegraphics[clip,width=0.48\textwidth]{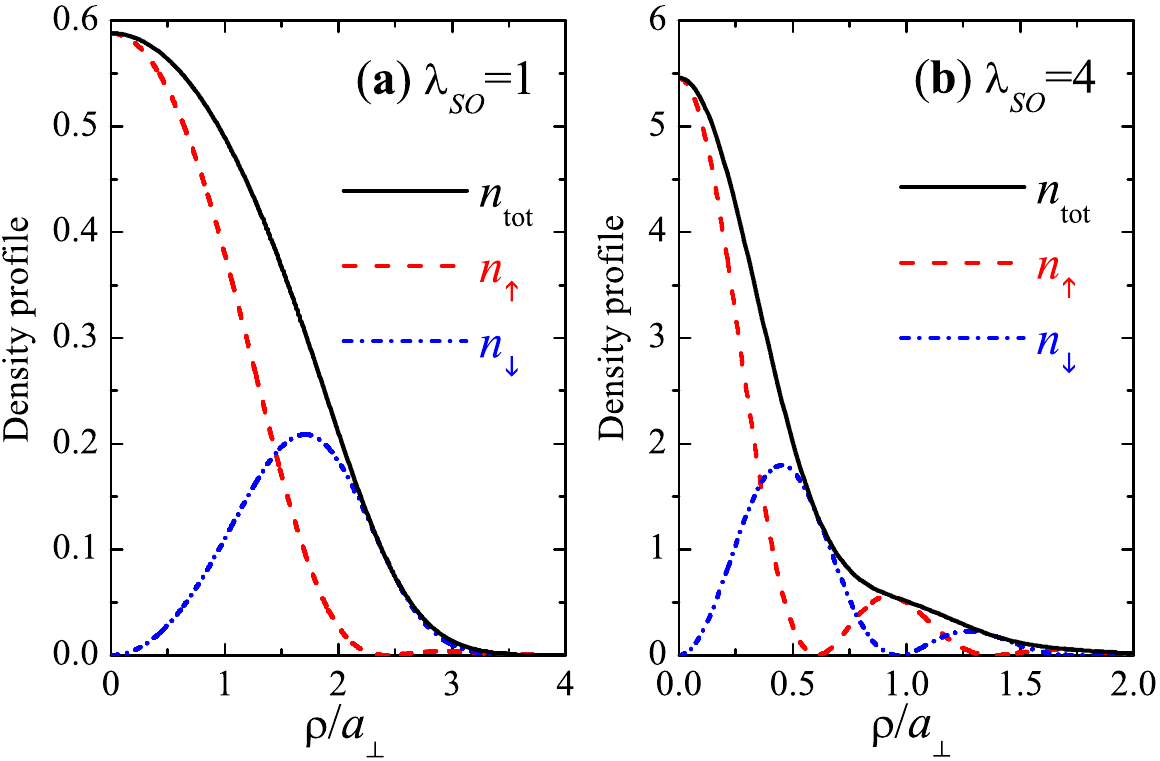} 
\par\end{centering}

\caption{(color online). Density distributions at $\lambda_{SO}=1$ and $g(N-1)=40\hbar\omega_{\perp}/a_{\perp}^{2}$
(a) and at $\lambda_{SO}=4$ and $g(N-1)=\hbar\omega_{\perp}/a_{\perp}^{2}$
(b). Here, the ratio $g_{\uparrow\downarrow}/g=1.1$.}

\label{fig3} 
\end{figure}

In Fig.~\ref{fig3}, we present the radial density distributions
of the half-quantum vortex condensate state at two SO coupling strengths:
$\lambda_{SO}=1$ and $\lambda_{SO}=4$. The increase of the SO coupling
leads to more oscillations in the radial direction. By comparing Fig.~\ref{fig3}(a)
with Fig.~\ref{fig2}(b), one finds that the density distributions
are flattened significantly by interatomic interactions, as anticipated.
The 2D contour plot of the spin-up and spin-down density patterns
of the half-quantum vortex state is shown in the inset of Fig.~\ref{fig1}
(in the phase I).

To gain more insights of the half-quantum vortex state, it is useful
to calculate the spin vector 
\begin{equation}
{\bf S}\left({\bf r}\right)=\frac{1}{2}\frac{\Phi^{\dagger}\sigma\Phi}{\left|\Phi\right|^{2}}
\end{equation}
 and the skyrmion density 
\begin{equation}
n_{{\rm skyrmion}}\left({\bf r}\right)=\frac{8}{4\pi}{\bf S}\cdot\left[\partial_{x}{\bf S}\times\partial_{y}{\bf S}\right].
\end{equation}
 The skyrmion density is a measure of the winding of the spin profile.
If it integrates to $1$ or $-1$, a topological stable knot exists
in the spin texture.

\begin{figure}[htp]
\begin{centering}
\includegraphics[clip,width=0.48\textwidth]{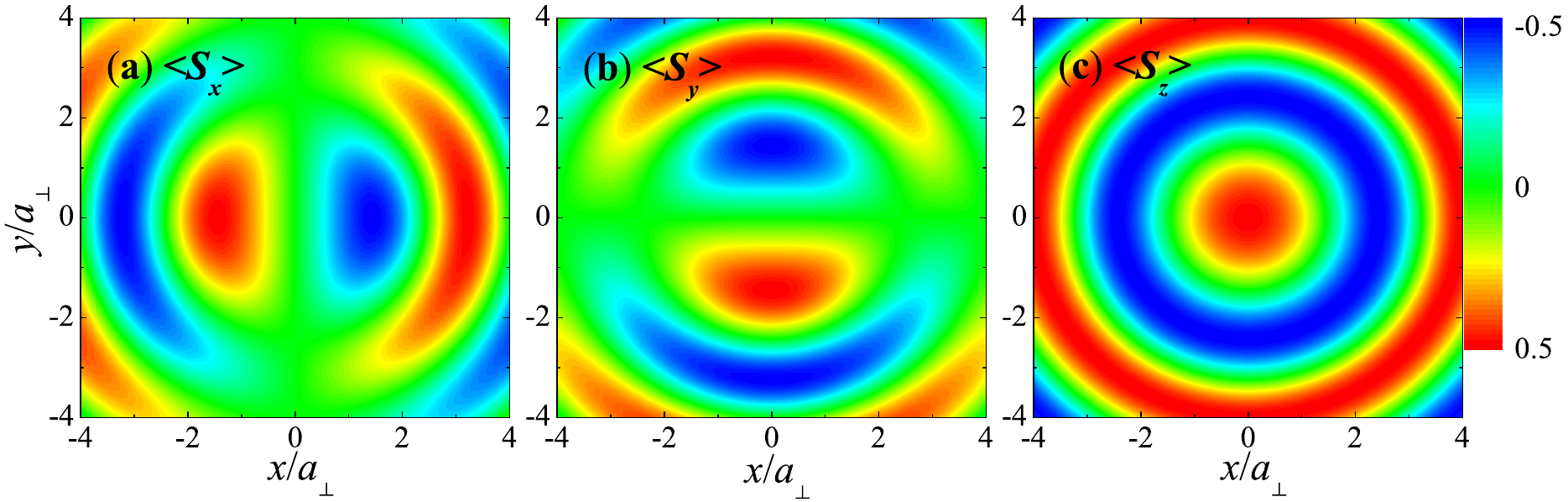} 
\par\end{centering}

\caption{(color online). Contour plots of the three components of spin vector
${\bf S}\left({\bf r}\right)$ at $\lambda_{SO}=1$, $g(N-1)=40\hbar\omega_{\perp}/a_{\perp}^{2}$
and $g_{\uparrow\downarrow}/g=1.1$.}

\label{fig4} 
\end{figure}

\begin{figure}[htp]
\begin{centering}
\includegraphics[clip,width=0.48\textwidth]{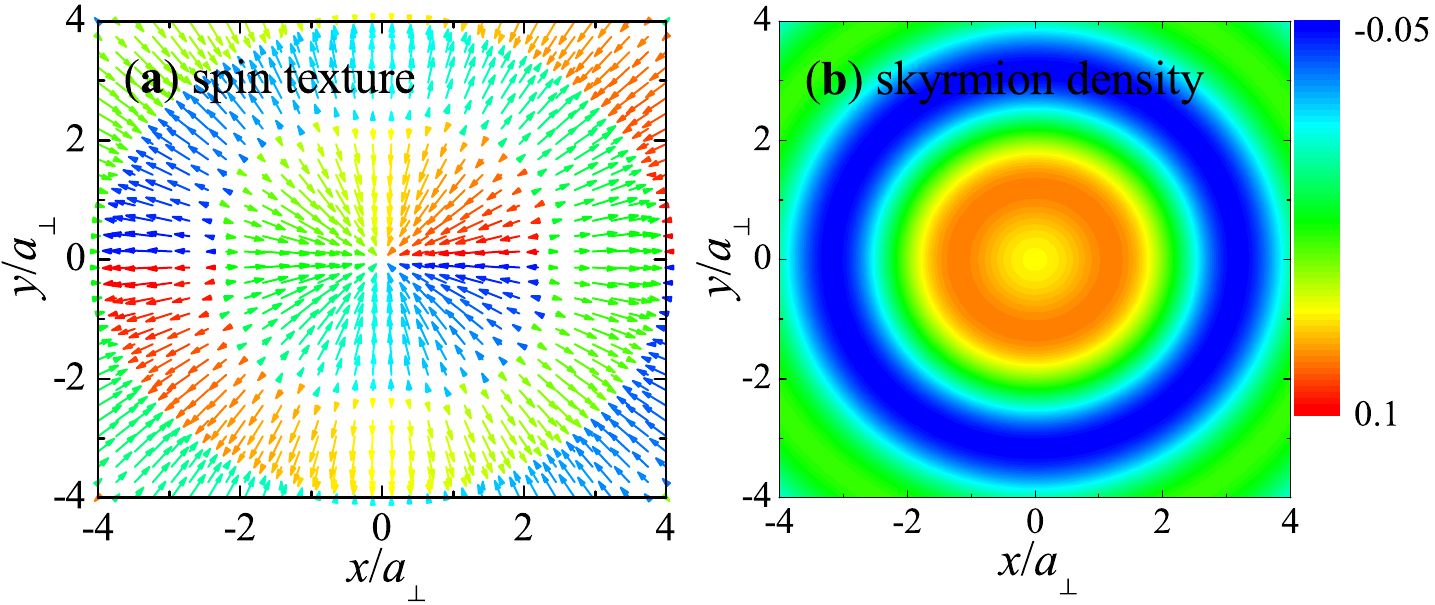} 
\par\end{centering}

\caption{(color online) (a) Two-dimensional vector plot of the transverse spin
vector $\left(S_{x,}S_{y}\right)$ at $\lambda_{SO}=1$, $g(N-1)=40\hbar\omega_{\perp}/a_{\perp}^{2}$
and $g_{\uparrow\downarrow}/g=1.1$. The color and length of arrows
give respectively the orientation and the magnitude of $\left(S_{x,}S_{y}\right)$.
(b) The corresponding skyrmion density $n_{{\rm skyrmion}}\left({\bf r}\right)$.}

\label{fig5} 
\end{figure}

In Fig.~\ref{fig4}, we report the three components of the spin vector
at $\lambda_{SO}=1$, $g(N-1)=40\hbar\omega_{\perp}/a_{\perp}^{2}$
and $g_{\uparrow\downarrow}/g=1.1$. The transverse spin texture is
shown in Fig. 5a by arrows, with color and length representing the
orientation and the magnitude of the transverse spin vector $\left(S_{x,}S_{y}\right)$,
respectively. It is readily seen that the spin vector spirals in space
and form a skyrmion-type texture. Quantitatively, this is most clearly
illustrated in Fig. 5b, where we plot the skyrmion density.

\subsection{Solutions of Bogoliubov equations}

Given the wave-function of the half-quantum vortex state, $[\phi_{\uparrow}\left(\rho\right),\phi_{\downarrow}\left(\rho\right)e^{i\varphi}]^{T}/\sqrt{2\pi}$,
we now turn to consider its collective excitations, as described by
the coupled Bogoliubov equations (\ref{Bog}). As a result of rotational
symmetry, it is easy to see that, the Bogoliubov wave-functions have
a good azimuthal quantum number $m$ and can be written as, $[u_{\uparrow}\left(\rho\right),u_{\downarrow}\left(\rho\right)e^{i\varphi},v_{\uparrow}\left(\rho\right),v_{\downarrow}\left(\rho\right)e^{-i\varphi}]^{T}e^{im\varphi}/\sqrt{2\pi}$.
Therefore, we have 
\begin{equation}
{\cal H}_{{\rm Bog}}\left[\begin{array}{c}
u_{\uparrow}\left(\rho\right)\\
u_{\downarrow}\left(\rho\right)\\
v_{\uparrow}\left(\rho\right)\\
v_{\downarrow}\left(\rho\right)
\end{array}\right]=\hbar\omega\left[\begin{array}{c}
+u_{\uparrow}\left(\rho\right)\\
+u_{\downarrow}\left(\rho\right)\\
-v_{\uparrow}\left(\rho\right)\\
-v_{\downarrow}\left(\rho\right)
\end{array}\right],\label{Bog2}
\end{equation}
 where 
\begin{equation}
{\cal H}_{{\rm Bog}}=\left[\begin{array}{cc}
{\cal L}_{m}+{\cal U} & {\cal U}\\
{\cal U} & {\cal L}_{-m}+{\cal U}
\end{array}\right],
\end{equation}
 with 
\begin{equation}
{\cal L}_{m}=\left[\begin{array}{cc}
{\cal H}_{s,m}+\bar{g}\phi_{\uparrow}^{2}+\bar{g}_{\uparrow\downarrow}\phi_{\downarrow}^{2} & \lambda_{R}\left[\partial_{\rho}+(m+1)/\rho\right]\\
\lambda_{R}\left(-\partial_{\rho}+m/\rho\right) & {\cal H}_{s,m+1}+\bar{g}_{\uparrow\downarrow}\phi_{\uparrow}^{2}+\bar{g}\phi_{\downarrow}^{2}
\end{array}\right],
\end{equation}
 and 
\begin{equation}
{\cal U}=\left[\begin{array}{cc}
\bar{g}\phi_{\uparrow}^{2} & \bar{g}_{\uparrow\downarrow}\phi_{\uparrow}\phi_{\downarrow}\\
\bar{g}_{\uparrow\downarrow}\phi_{\uparrow}\phi_{\downarrow} & \bar{g}\phi_{\downarrow}^{2}
\end{array}\right].
\end{equation}

To solve the Bogoliubov equation, as before we expand the wave-functions
using 2D harmonic oscillator basis, 
\begin{eqnarray}
u_{\uparrow}\left(\rho\right) & = & \sum_{k}a_{k}R_{km}\left(\rho\right),\\
u_{\downarrow}\left(\rho\right) & = & \sum_{k}b_{k}R_{km+1}\left(\rho\right),\\
v_{\uparrow}\left(\rho\right) & = & \sum_{k}c_{k}R_{km}\left(\rho\right),\\
v_{\downarrow}\left(\rho\right) & = & \sum_{k}d_{k}R_{km-1}\left(\rho\right).
\end{eqnarray}
 This leads to a secular matrix of ${\cal H}_{{\rm Bog}}$, whose
elements can be calculated directly using the 2D harmonic oscillator
basis. We note that, to obtain the Bogoliubov quasiparticles we cannot
diagonalize directly the secular matrix, because of the minus sign
before $v_{\uparrow}\left(\rho\right)$ and $v_{\downarrow}\left(\rho\right)$
at the right-hand side of Eq. (\ref{Bog2}). Instead, we should diagonalize
a {\em non-symmetric} matrix Diag$\{+1,+1,-1,-1\}{\cal H}_{{\rm Bog}}$
and normalize the quasi-particle wave-functions according to $\int_{0}^{\infty}\rho d\rho[u_{\uparrow}^{2}+u_{\downarrow}^{2}-v_{\uparrow}^{2}-v_{\downarrow}^{2}]=1$.
The number of resulting eigenvalues is two times the number that we
want. There are two branches of eigenvalues, one is positive and the
other negative, as a result of the duality between the solution $[u_{\uparrow}\left({\bf r}\right),u_{\downarrow}\left({\bf r}\right),v_{\uparrow}\left({\bf r}\right),v_{\downarrow}\left({\bf r}\right)]^{T}$
(with energy $+\hbar\omega$) and $[v_{\uparrow}^{*}\left({\bf r}\right),v_{\downarrow}^{*}\left({\bf r}\right),u_{\uparrow}^{*}\left({\bf r}\right),u_{\downarrow}^{*}\left({\bf r}\right)]^{T}$
(with energy $-\hbar\omega$). We should take the positive branch.
We note also that the Bogoliubov quasi-particles at a negative azimuthal
quantum number $m$ may be obtained from the negative branch of the
solution with $m>0$, because of the duality.

\subsubsection{Breathing modes}

In the case of the breathing mode ($m=0$), where 
\begin{equation}
{\cal H}_{{\rm Bog}}=\left[\begin{array}{cc}
{\cal L}_{{\rm GP}}+{\cal U} & {\cal U}\\
{\cal U} & {\cal L}_{{\rm GP}}+{\cal U}
\end{array}\right],
\end{equation}
 we may have an alternative way to solve the Bogoliubov equation,
following Hutchinson, Zaremba, and Griffin (HZG) \cite{HZG}. By denoting
collectively $u=[u_{\uparrow}\left(\rho\right),u_{\downarrow}\left(\rho\right)]$
and $v=[v_{\uparrow}\left(\rho\right),v_{\downarrow}\left(\rho\right)]$,
we have, 
\begin{eqnarray}
\left({\cal L}_{{\rm GP}}+2{\cal U}\right)\left(u+v\right) & = & \hbar\omega\left(u-v\right),\label{Bog3a}\\
{\cal L}_{{\rm GP}}\left(u-v\right) & = & \hbar\omega\left(u+v\right).\label{Bog3b}
\end{eqnarray}
 Let us now expand the wave-functions $u\pm v$ in terms of the eigenfunctions
$\psi_{\alpha}$ of ${\cal L}_{{\rm GP}}$ with energy $\epsilon_{\alpha}$
(i.e., ${\cal L}_{{\rm GP}}\psi_{\alpha}=\epsilon_{\alpha}\psi_{\alpha}$),
\begin{eqnarray}
u-v & = & \sum_{\alpha\neq0}\frac{c_{\alpha}}{\epsilon_{\alpha}^{1/2}}\psi_{\alpha},\\
u+v & = & \sum_{\alpha\neq0}\frac{\epsilon_{\alpha}^{1/2}c_{\alpha}}{\hbar\omega}\psi_{\alpha}.
\end{eqnarray}
 Here, the lowest eigenstate of ${\cal L}_{{\rm GP}}$ with zero energy
should be removed, as it corresponds exactly to the condensate mode.
It is easy to see that $({\cal L}_{{\rm GP}}+2{\cal U}){\cal L}_{GP}(u-v)=(\hbar\omega)^{2}(u-v)$
and ${\cal L}_{{\rm GP}}({\cal L}_{{\rm GP}}+2{\cal U})(u+v)=(\hbar\omega)^{2}(u+v)$.
Inserting the expansion of $u-v$ or $u+v$, one finds the secular
equation, 
\begin{equation}
\sum_{\beta}\left\{ \epsilon_{\alpha}^{2}\delta_{\alpha\beta}+2\epsilon_{\alpha}^{1/2}{\cal U}_{\alpha\beta}\epsilon_{\beta}^{1/2}\right\} c_{\beta}=\left(\hbar\omega\right)^{2}c_{\alpha},
\end{equation}
 where 
\begin{equation}
{\cal U}_{\alpha\beta}=\int_{0}^{\infty}\rho d\rho\;\psi_{\alpha}^{\dagger}\left(\rho\right){\cal U}\psi_{\beta}\left(\rho\right).
\end{equation}
 By diagonalizing the secular matrix, one obtains the mode frequency
$\omega$ and the coefficients $c_{\alpha}$. The latter should be
normalized as $\sum_{\alpha}c_{\alpha}^{2}=\hbar\omega$, in accord
with the normalization condition for $u$ and $v$.

We have checked numerically that the HZG solution leads to exactly
the same result as the direct diagonalization of the non-symmetric
matrix Diag$\{+1,+1,-1,-1\}{\cal H}_{{\rm Bog}}$, if we discard the
zero-frequency condensate mode in the latter method.

\subsection{Collective excitations}

\begin{figure}[htp]
\begin{centering}
\includegraphics[clip,width=0.48\textwidth]{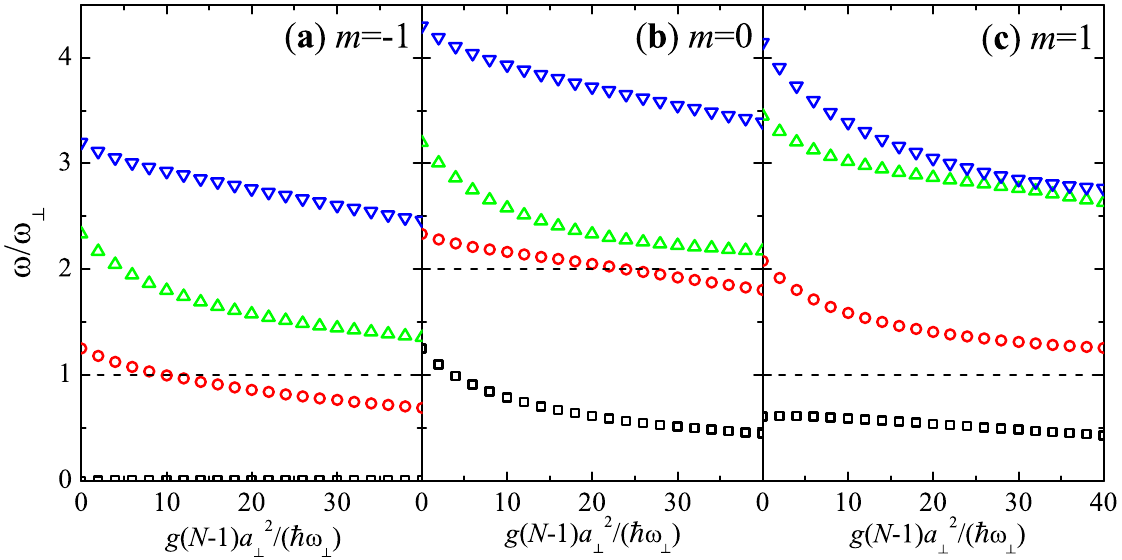} 
\par\end{centering}

\caption{(color online). The mode frequency of breathing ($m=0$) and dipole
($m=\pm1$) modes as a function of interaction strength at a fixed
SO coupling $\lambda_{SO}=1$ and at $g_{\uparrow\downarrow}=1.1g$.}

\label{fig6} 
\end{figure}

\begin{figure}[htp]
\begin{centering}
\includegraphics[clip,width=0.48\textwidth]{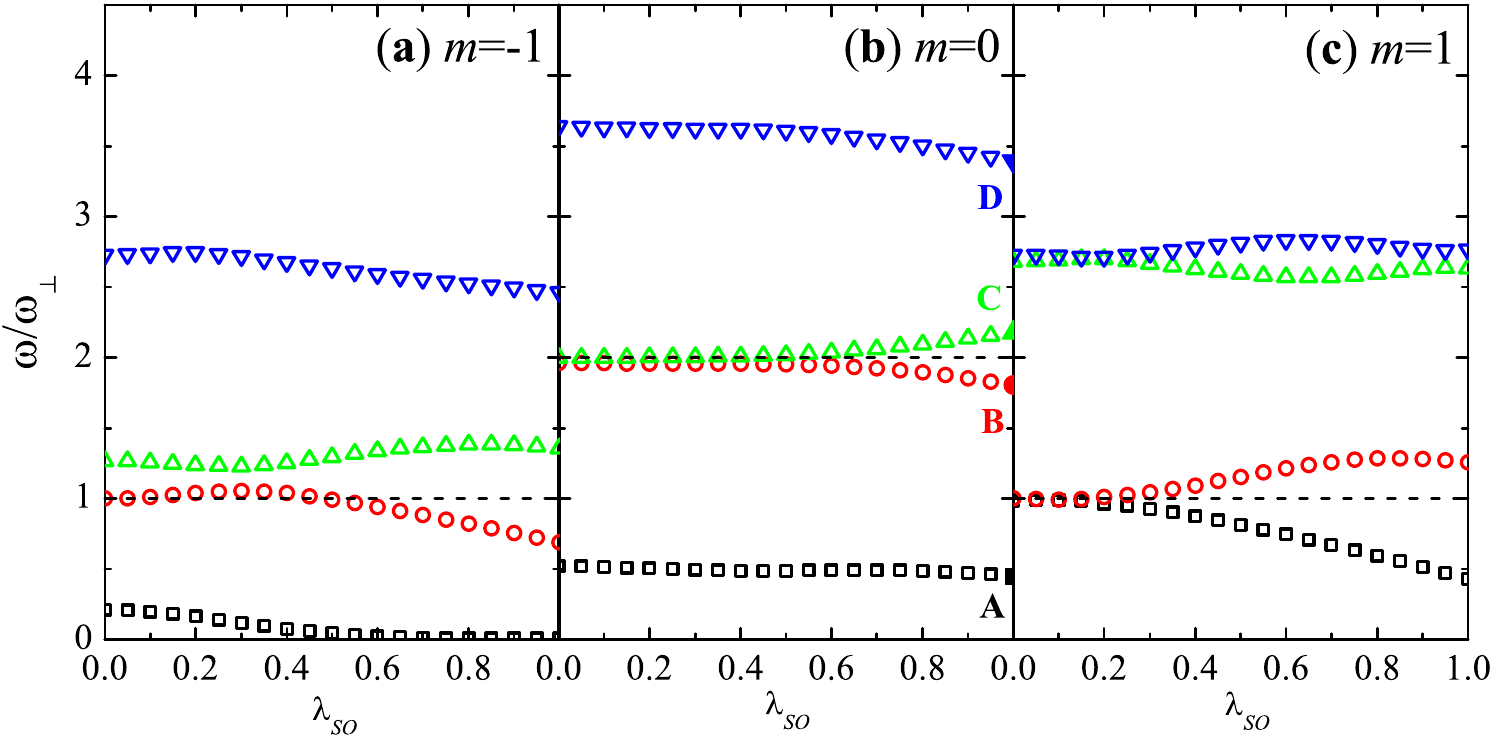} 
\par\end{centering}

\caption{(color online). The mode frequency of breathing ($m=0$) and dipole
($m=\pm1$) modes as a function of SO coupling at a fixed interaction
strength $g(N-1)=40\hbar\omega_{\perp}/a_{\perp}^{2}$ and at $g_{\uparrow\downarrow}=1.1g$.}

\label{fig7} 
\end{figure}

In Fig. 6, we report the breathing ($m=0$) and the dipole mode ($m=\pm1$)
frequencies as a function of the interaction strength. With increasing
interaction, the mode frequency decreases and seems to saturate at
sufficiently large interactions. This may be anticipated from the
point of view of two-fluid hydrodynamic behavior in the Thomas-Fermi
regime. In Fig. 7, we report the dependence of the mode frequencies
on SO coupling. In the absence of SO coupling, the breathing mode
with $\omega=2\omega_{\perp}$ and the dipole mode with $\omega=\omega_{\perp}$
are the exact solutions of quantum many-body systems in harmonic traps.
At a finite SO coupling, however, we find that these two solutions
are no longer exact. The relative deviations of the breathing mode
and dipole mode at $\lambda_{SO}=1$ are about $10\%$ and $30\%$,
respectively.

\begin{figure}[htp]
\begin{centering}
\includegraphics[clip,width=0.48\textwidth]{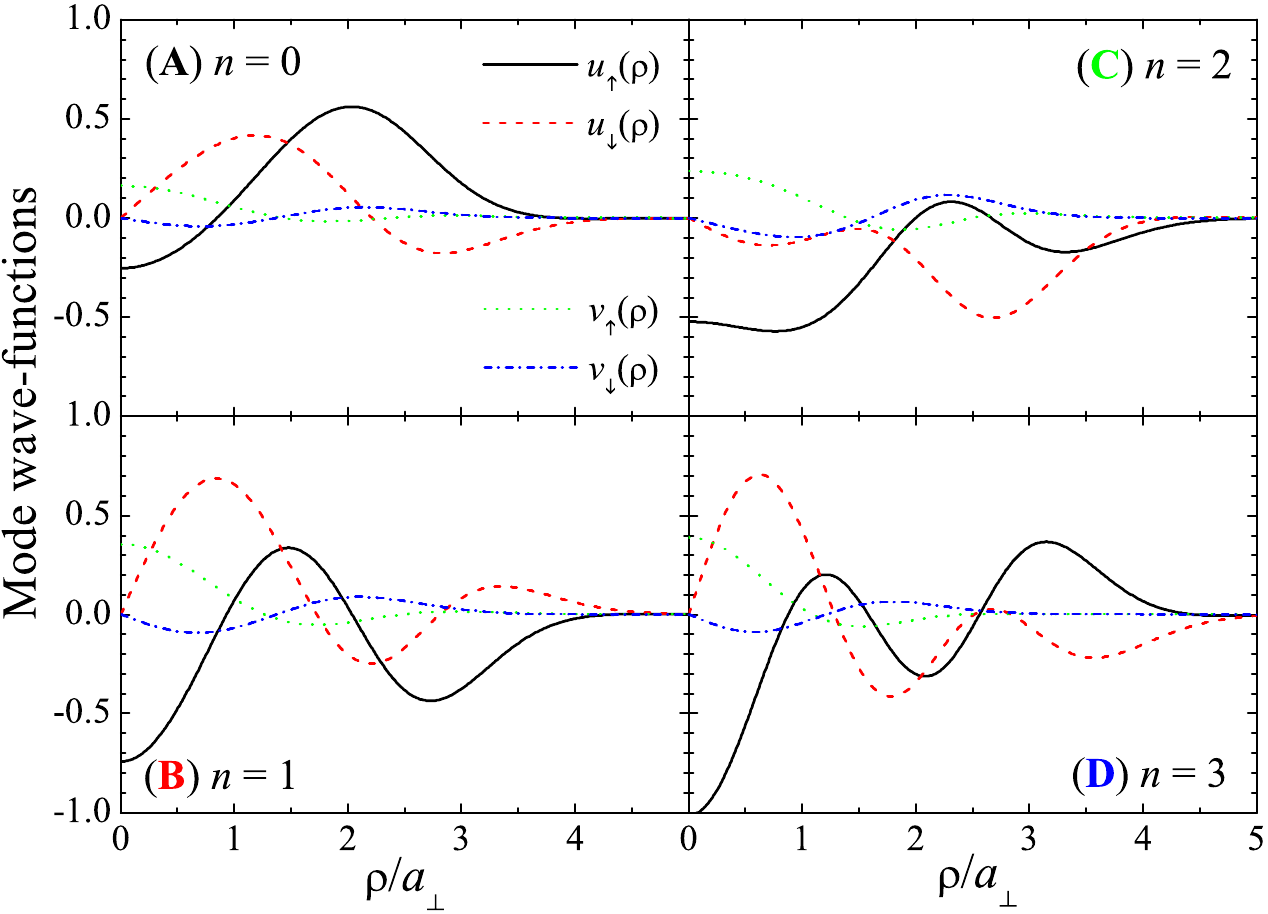} 
\par\end{centering}

\caption{(color online). Bogoliubov wave-functions of the lowest four breathing
modes at $\lambda_{SO}=1$, $g(N-1)=40\hbar\omega_{\perp}/a_{\perp}^{2}$
and $g_{\uparrow\downarrow}=1.1g$. The mode frequencies are indicated
in Fig. 7b by solid symbols.}

\label{fig8} 
\end{figure}

In Fig. 8, we plot the Bogoliubov wave-functions of the lowest four
breathing modes at $\lambda_{SO}=1$, $g(N-1)=40\hbar\omega_{\perp}/a_{\perp}^{2}$
and $g_{\uparrow\downarrow}=1.1g$. We find that the density response
is mainly carried by $u_{\uparrow}(\rho)$ and $u_{\downarrow}(\rho)$
components. With increasing mode frequency, more and more nodes appear
in $u_{\uparrow}(\rho)$ and $u_{\downarrow}(\rho)$. In contrast,
the response in $v_{\uparrow}(\rho)$ and $v_{\downarrow}(\rho)$
is relatively weak and the curve shape is nearly unchanged as the
mode frequency increases.


\subsection{Dynamical Calculations}

To investigate the dynamical properties of the system, we also perform
direct simulations of the system by real-time propagation of the ground
state under perturbation. To do this, firstly we obtain the ground
state by solving the coupled GP equations in Eqn.~(\ref{HGP}) using
the TSSP technique. The half-quantum vortex ground state is perturbed
in various ways. We observe that the mode frequencies obtained by
dynamical simulation agree well with those obtained by solving Bogoliubov
equations (shown in Fig.~\ref{fig6}). 

\begin{figure}[htp]
\begin{centering}
\includegraphics[clip,width=0.45\textwidth]{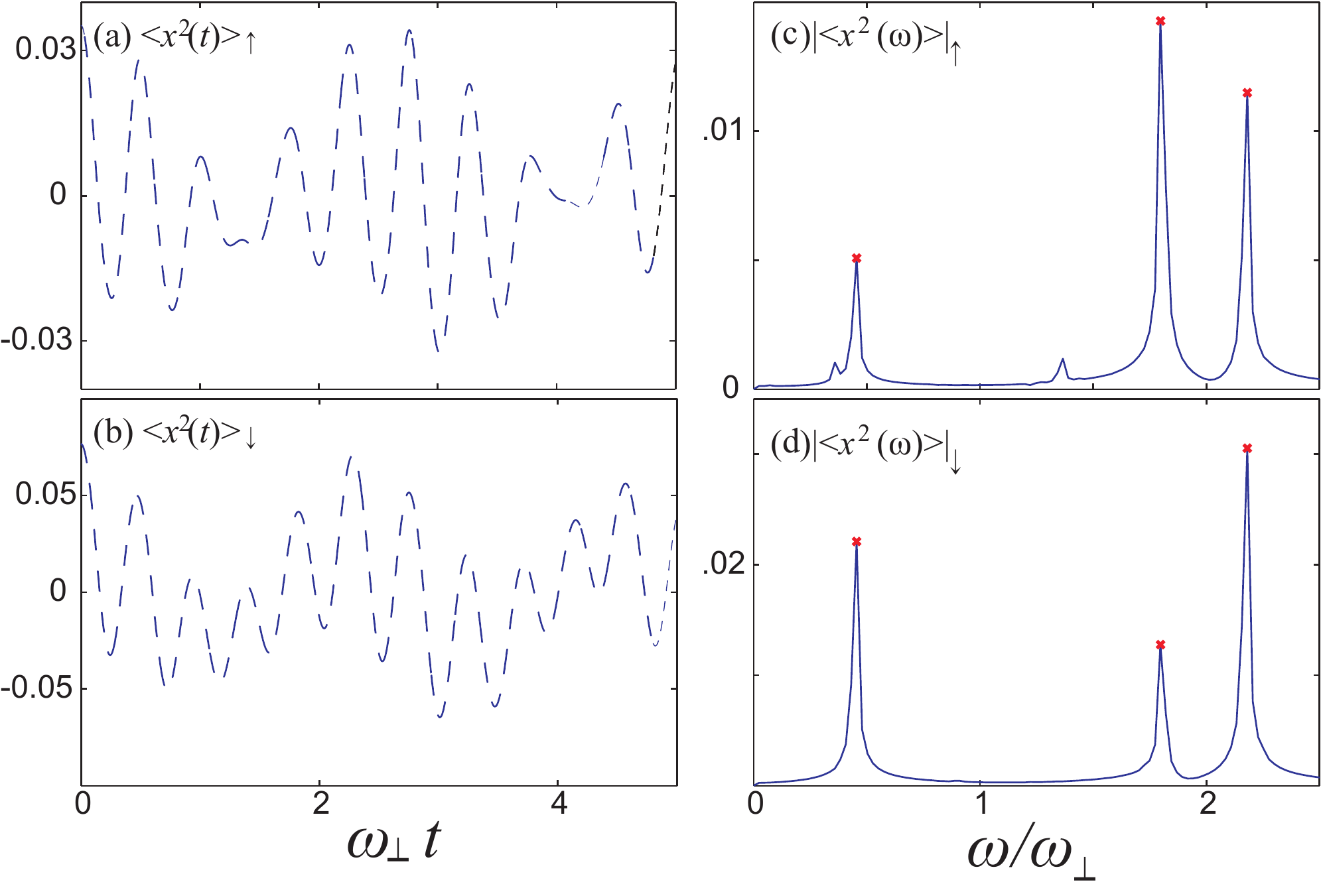} 
\par\end{centering}

\caption{(color online). (a),(b): Dynamic response of the mean square of the
center-of-mass coordinate in $x$-direction of $\uparrow$- and $\downarrow$-
spin components respectively. We have shifted the curves by subtracting
the time-averaged $\langle x^{2}(t)\rangle_{\sigma}$. Without this
shift, the Fourier spectrum as shown in (c) and (d) will be dominated
by a large peak at $\omega=0$. (c),(d): Corresponding single-sided
amplitude spectrum of the collective coordinate. Parameters used:
$\lambda_{SO}=1.0,g(N-1)=40\hbar\omega_{\perp}/a_{\perp}^{2},g_{\uparrow\,\downarrow}/g=1.1$.}

\label{figmono} 
\end{figure}

\emph{ Breathing mode analysis, $m=0$}: We excite the monopole mode
by weak relaxation of the trapping frequency at time $t=0$, and letting
the system propagate in real-time. As the breathing mode excitation
is isotropic in $x$-$y$ space, it is sufficient to observe the dynamic
response of the collective coordinate along one axis, say, the $x$-axis.
Here, we pick the mean square of the center-of-mass coordinate as
the quantity of interest: 
\[
\langle x^{2}\rangle_{\sigma}=\frac{\int{|\phi_{\sigma}|^{2}x^{2}dx\, dy}}{\int{|\phi_{\sigma}|^{2}dx\, dy}}\,,
\]
 where $\sigma={\uparrow,\downarrow}$-spin components. In Fig.~\ref{figmono}
(a),(b), we plot the time response of $\langle x^{2}(t)\rangle_{\sigma}$
for a typical parameter set. In Fig.~\ref{figmono} (c),(d), we show
the corresponding frequency response by plotting the single-sided
amplitude spectrum $\vert\langle x^{2}(\omega)\rangle\vert_{\sigma}$,
which are just the Fourier transforms of $\langle x^{2}(t)\rangle_{\sigma}$.
We observe frequency peaks at $\omega/\omega_{\perp}\simeq0.46,1.8,2.18$
and at 3.40 (not shown). We note that these values exactly match with
the mode frequencies obtained for this parameter set by solving Bogoliubov
equations, shown in Fig.~\ref{fig6}(b).

\begin{figure}[htp]
\begin{centering}
\includegraphics[clip,width=0.45\textwidth]{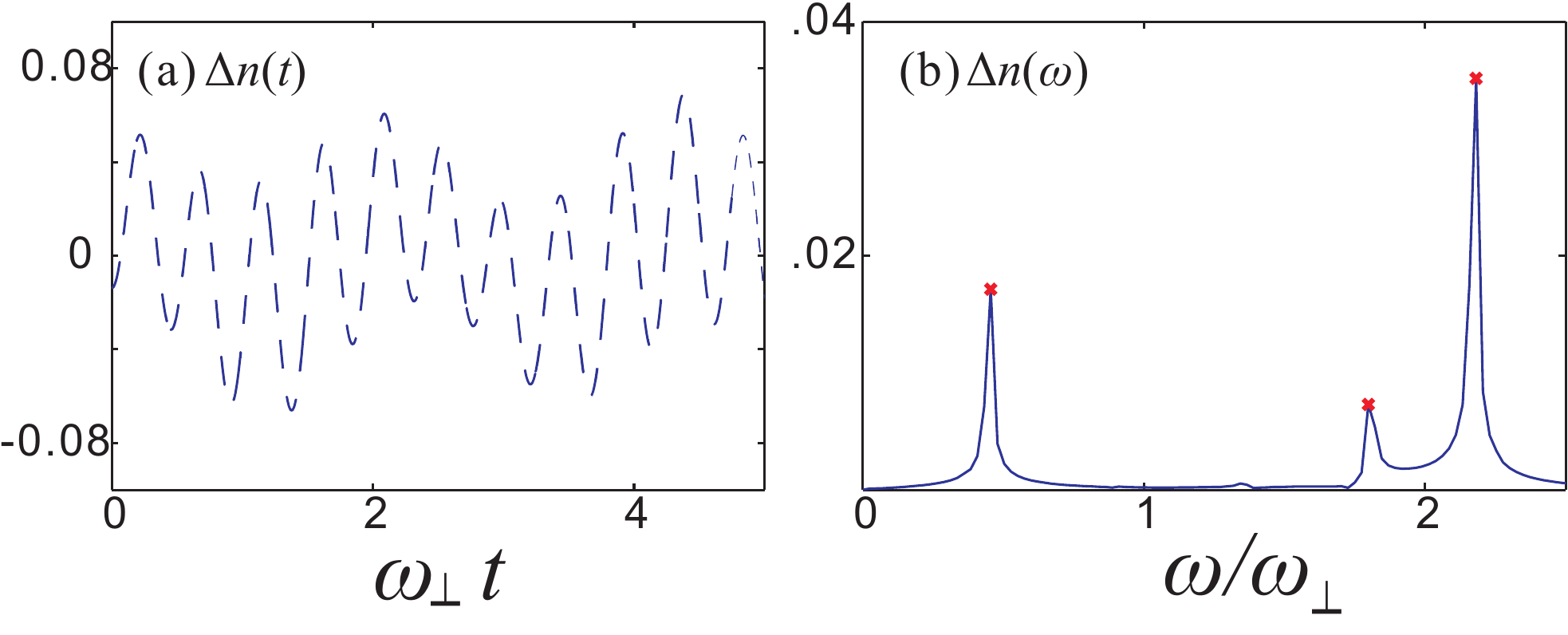} 
\par\end{centering}

\caption{(color online). (a) Dynamic response and (b) single-sided amplitude
spectrum of population difference $\Delta n$ for the same parameter
set used in Fig.~\ref{figmono}. }

\label{Szdyn} 
\end{figure}

Dynamical calculations also reveal the coupling between the center-of-mass
motion and the internal spin degrees of freedom, a trademark signature
of spin-orbit coupled systems. We shall now discuss the dynamic response
of the population difference $\Delta n=\int d{\bf r}\,(|\phi_{\uparrow}|^{2}-|\phi_{\downarrow}|^{2})$.
In Fig.~\ref{Szdyn}(a), we plot the time response of $\Delta n(t)$
for the same parameter set mentioned in Fig.~\ref{figmono}. In Fig.~\ref{Szdyn}(b),
we show the corresponding frequency response by plotting the single-sided
amplitude spectrum $\vert\Delta n(\omega)\vert$. We observe frequency
peaks at $\omega/\omega_{\perp}\simeq0.46,1.8,2.18$ and at 3.40 (not
shown), exactly matching with the mode frequencies obtained in Fig.~\ref{figmono}.
This analysis clearly shows that the population transfer between the
two spin components shares a similar dynamic response with the collective
motional coordinate. In this aspect, response of $\Delta n$ in a
spin-orbit coupled spinor BEC (shown here) is similar to the effects
observed in the presence of internal Josephson coupling in multi-component
condensates \cite{Ohberg1999}.

\begin{figure}[htp]
\begin{centering}
\includegraphics[clip,width=0.45\textwidth]{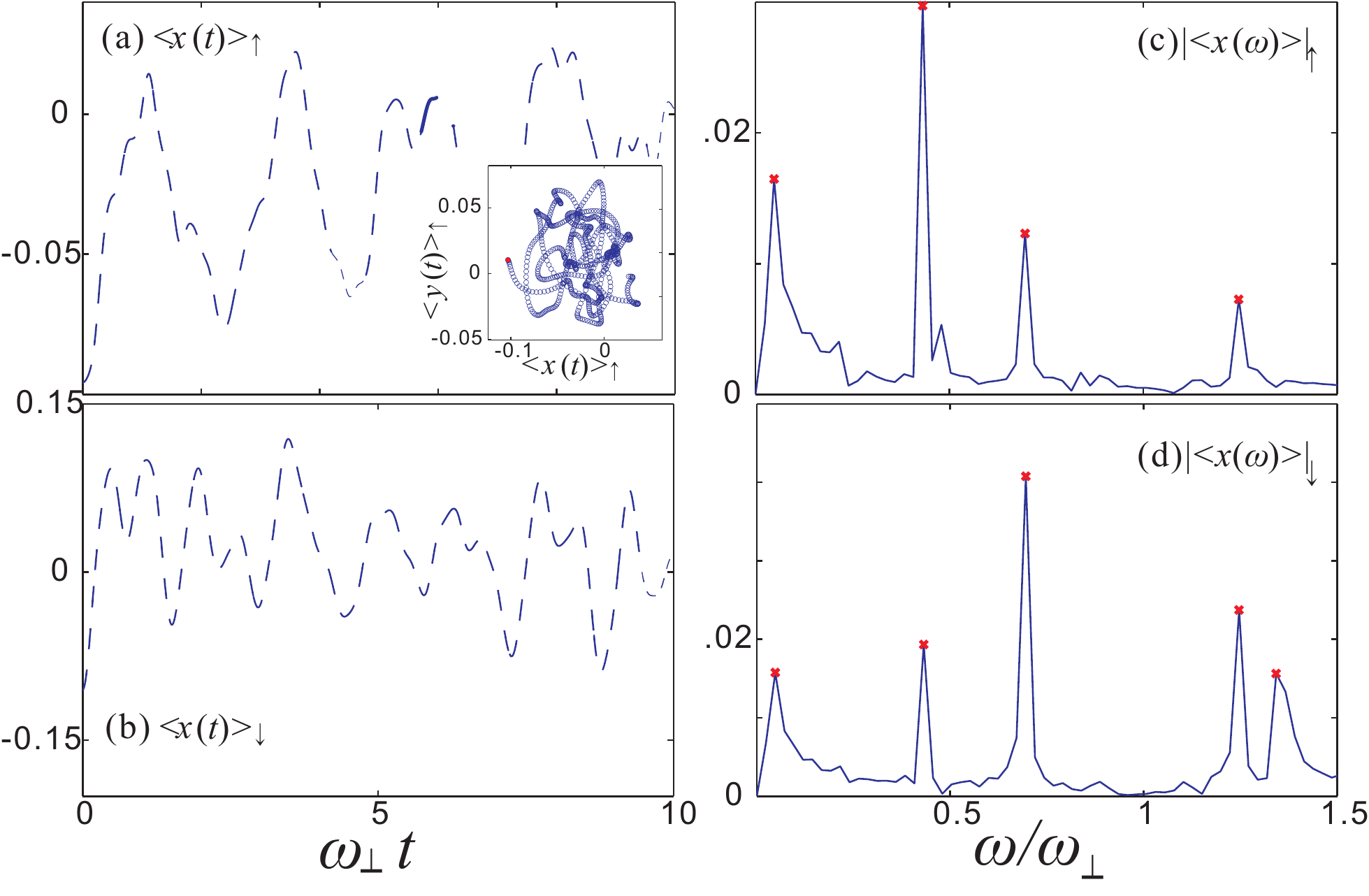} 
\par\end{centering}

\caption{(color online). Parameters used: $\lambda_{SO}=1.0,g(N-1)=40\hbar\omega_{\perp}/a_{\perp}^{2},g_{\uparrow\,\downarrow}=1.1\, g$.
(a),(b): Dynamic response of the center-of-mass coordinate in x-direction
of $\uparrow$- and $\downarrow$- spin components respectively. The
inset in (a) shows the dynamics of the center-of-mass coordinate over
12 trap periods and the filled (red) marker denotes the initial position.
(c),(d): Corresponding single-sided amplitude spectrum of the collective
coordinate.}

\label{figdi} 
\end{figure}

\emph{Dipole mode analysis, $m=\pm1$:} We excite the dipole modes
by displacing the trap in $x$-direction by a small amount at time
$t=0$, and letting the system propagate in real-time. We observe
the dynamic response of the center-of-mass coordinate in x-direction:
\[
\langle x\rangle_{\sigma}=\frac{\int{|\phi_{\sigma}|^{2}\, x\, dx\, dy}}{\int{|\phi_{\sigma}|^{2}dx\, dy}}\,.
\]
 In Fig.~\ref{figdi} (a),(b), we plot the time response of this
collective coordinate in $x$-direction of $\uparrow$- and $\downarrow$-
spin components for a typical parameter set. In Fig.~\ref{figdi}
(c),(d), we show the corresponding frequency response by plotting
the single-sided amplitude spectrum $\vert\langle x(\omega)\rangle\vert_{\sigma}$.
We observe frequency peaks at $\omega/\omega_{\perp}\simeq0.05,0.43,0.70,1.25,1.34,$
(shown) and at 2.5, 2.64, 2.76 (not shown). We note that these values
exactly agree the mode frequencies obtained for this parameter set
by solving Bogoliubov equations, shown in Fig.~\ref{fig6}(a),(c).

In the inset of Fig.~\ref{figdi}(a), we show the dynamics of the
center-of-mass coordinate. It is important to note that even though
the trap is displaced only in the $x$-direction, we also observe
a similar dynamic response in $y$-direction of both spin components
(only $\uparrow$-spin component shown). This behavior occurs due
to the vorticity induced by the spin-orbit coupling --- the vortex
state experiences a Magnus force that is perpendicular to its motion.
Hence a displacement in the $x$-direction induces a motion along
the $y$-direction. Furthermore, the trace of the center-of-mass and
its magnitude are affected by the strength of the inter-particle interactions
and the spin-orbit coupling induced population transfer, as observed
in the case of the breathing mode excitation, between the $\uparrow$-
and $\downarrow$- spin components.

\section{Instability analysis and phase diagram}

We are now ready to analyze the parameter space for the existence
of half-quantum vortex state. It will become unstable with respect
to increasing the interaction strength or decreasing the ratio $g_{\uparrow\downarrow}/g$.
The instability could be indicated from some energy considerations
and from the softening of collective density modes.

\subsection{Superposition instability}

\label{super} As we mentioned earlier, for any half-quantum vortex
state, $\phi({\bf r})=[\phi_{\uparrow}(\rho),\phi_{\downarrow}(\rho)e^{i\varphi}]^{T}/\sqrt{2\pi}$,
we always have a degenerate time-reversal partner state, ${\cal T}\phi({\bf r})=[\phi_{\downarrow}(\rho)e^{-i\varphi},-\phi_{\uparrow}(\rho)]^{T}/\sqrt{2\pi}$.
There is an instability for half-quantum vortex state with respect
to a superposition state, which with equal weight takes the form,
\begin{equation}
\phi_{s}({\bf r})=\frac{1}{\sqrt{4\pi}}\left[\begin{array}{c}
\phi_{\uparrow}(\rho)+\phi_{\downarrow}(\rho)e^{-i\left(\varphi-\varphi_{0}\right)}\\
\phi_{\downarrow}(\rho)e^{i\left(\varphi-\varphi_{0}\right)}-\phi_{\uparrow}(\rho)
\end{array}\right].
\end{equation}
 Here $\varphi_{0}$ is an arbitrary azimuthal angle. The energy difference
between the superposition state and the half-quantum vortex state
is given by, 
\begin{equation}
\Delta E_{{\rm GP}}=\frac{\left(g_{\uparrow\downarrow}-g\right)(N-1)}{4}W[\phi({\bf r})].
\end{equation}
 Therefore, if $W[\phi({\bf r})]>0$, the half-quantum vortex state
is stable only when $g<g_{\uparrow\downarrow}$.

\begin{figure}[htp]
\begin{centering}
\includegraphics[clip,width=0.48\textwidth]{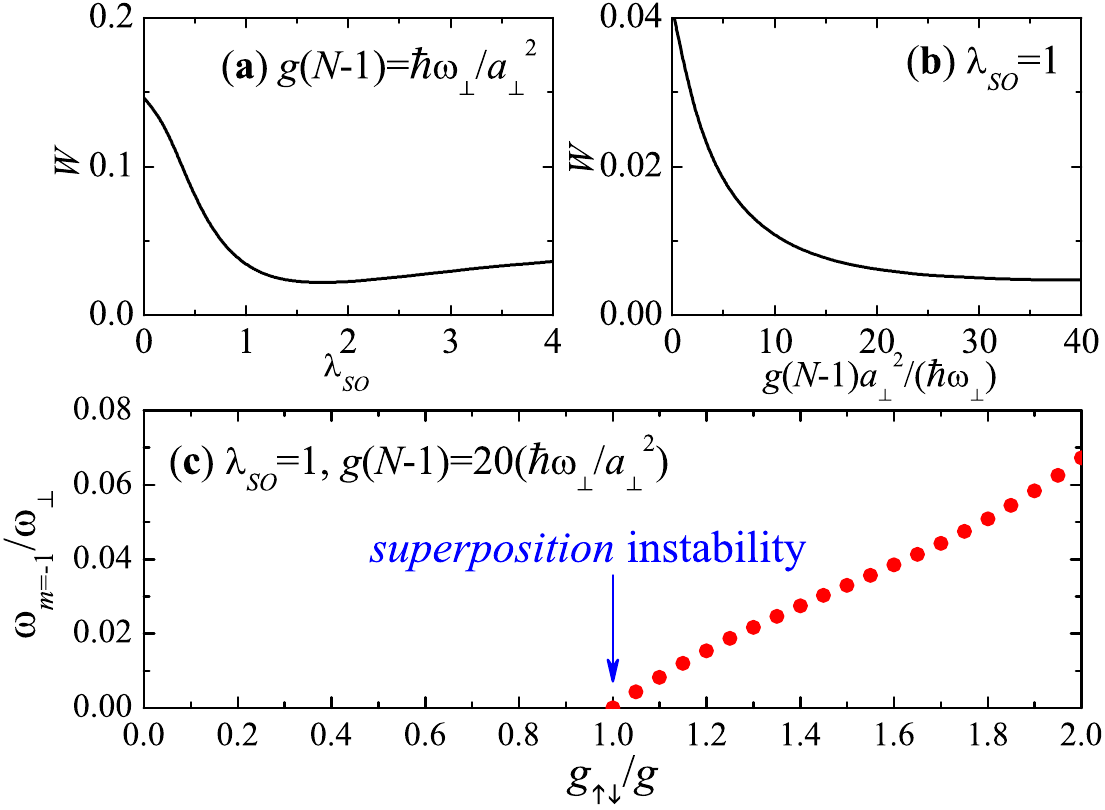} 
\par\end{centering}

\caption{(color online). (a) The $W$-function as a function of SO coupling
at $g(N-1)=\hbar\omega_{\perp}/a_{\perp}^{2}$ and $g_{\uparrow\downarrow}=1.1g$.
(b) The $W$-function as a function of interaction strength at $\lambda_{SO}=1$
and $g_{\uparrow\downarrow}=1.1g$. (c) The instability of the lowest
dipole mode frequency $\omega_{m=-1}$ with decreasing $g_{\uparrow\downarrow}/g$
at $\lambda_{SO}=1$ and $g(N-1)=20\hbar\omega_{\perp}/a_{\perp}^{2}$.}

\label{fig9} 
\end{figure}

In Figs.~\ref{fig9}(a) and (b), we check the $W$-function of the
half-quantum vortex state in the presence of interatomic interactions.
It always appears to be positive, though the interactions tend to
decrease its absolute magnitude. Hence, there must be a quantum phase
transition occurring at the isotropic point $g=g_{\uparrow\downarrow}$.
Once $g>g_{\uparrow\downarrow}$, a superposition state with density
pattern, 
\begin{equation}
n_{\uparrow,\downarrow}=\frac{1}{2\pi}\left[\frac{\phi_{\uparrow}^{2}+\phi_{\downarrow}^{2}}{2}\pm\phi_{\uparrow}\phi_{\downarrow}\cos\left(\varphi-\varphi_{0}\right)\right],
\end{equation}
 becomes preferable. The 2D contour plot of this density pattern with
$\varphi_{0}=0$ is schematically shown in the inset of Fig. 1 (in
the phase IIA).

In general, in passing the quantum phase transition point, we would
observe softening of a particular mode frequency. As the superposition
state involves a time-reversal state with angular momentum $m=-1$,
the lowest dipole mode with $m=-1$ may become unstable. In Fig.~\ref{fig9}(c),
we plot the lowest dipole mode frequency $\omega_{m=-1}$ as a function
of $g_{\uparrow\downarrow}/g$ at $\lambda_{SO}=1$ and $g(N-1)=20\hbar\omega_{\perp}/a_{\perp}^{2}$.
Indeed, with decreasing $g_{\uparrow\downarrow}/g$, the mode frequency
$\omega_{m=-1}$ decreases and approaches to zero exactly at the phase
transition point.

\subsection{Instability to high-order angular momentum components}

\label{highang} There is another instability for the half-quantum
vortex state, occurring with increasing the interatomic interactions.
With sufficiently large interactions, we anticipate that the state
with high-order azimuthal angular momentum will energetically become
favorable. For example, let us consider a condensate state with an
azimuthal angular momentum $m=1$ (the 3/2-quantum vortex state),
which has the form, 
\begin{equation}
\phi_{m=1}({\bf r})=\frac{1}{\sqrt{2\pi}}\left[\begin{array}{c}
\phi_{\uparrow}(\rho)e^{i\varphi}\\
\phi_{\downarrow}(\rho)e^{i2\varphi}
\end{array}\right].
\end{equation}
 The GP energy of this state can be obtained by solving the GPE equation
as before, except that we need to take $R_{k1}\left(\rho\right)$
and $R_{k2}\left(\rho\right)$ as the expansion functions for $\phi_{\uparrow}(\rho)$
and $\phi_{\downarrow}(\rho)$, respectively. Its degenerate time-reversal
partner state has an azimuthal angular momentum $m=-2$.

\begin{figure}[htp]
\begin{centering}
\includegraphics[clip,width=0.45\textwidth]{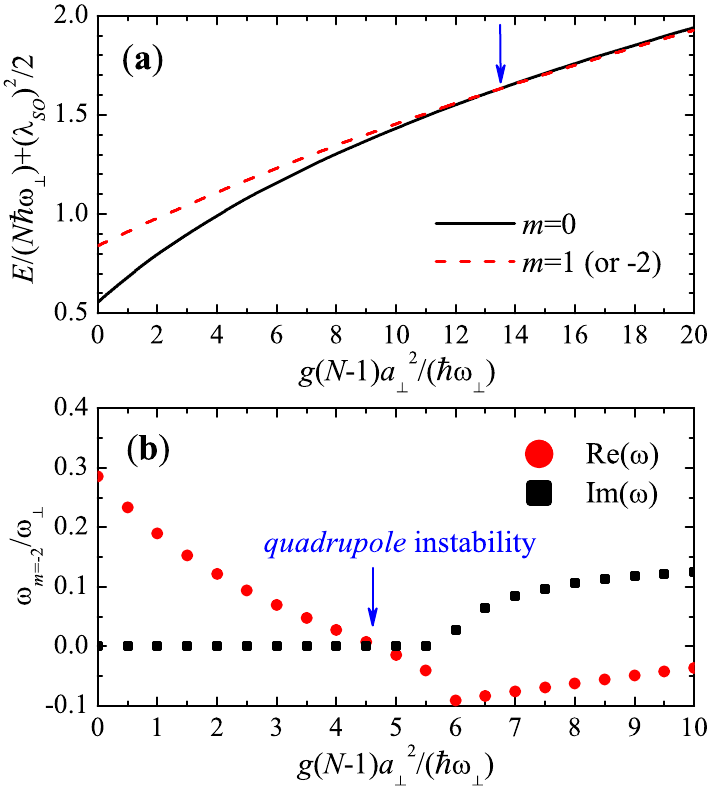} 
\par\end{centering}

\caption{(color online). (a) GP energy of the 3/2-quantum vortex state $\phi_{m=1}({\bf r})$
and of the half-quantum vortex state $\phi_{m=0}({\bf r})$ as a function
of interaction strength at $\lambda_{SO}=2$ and $g_{\uparrow\downarrow}/g=1.1$.
Beyond a critical interaction strength as indicated by an arrow, $\phi_{m=1}({\bf r})$
becomes energetically favorable. (b) The corresponding lowest quadrupole
mode frequency $\omega_{m=-2}$. It becomes unstable beyond a threshold
$g_{c}$.}

\label{fig10} 
\end{figure}

It is easy to see from Fig.~\ref{fig10}(a) that beyond a critical
interaction strength the condensate state with $m=1$, $\phi_{m=1}({\bf r})$,
is lower in energy than the half-quantum vortex state, $\phi_{m=0}({\bf r})$.
We note, however, that the critical interaction strength determined
in this way is not accurate, as a superposition state of $\phi_{m=0}({\bf r})$
and $\phi_{m=1}({\bf r})$ may already become energetically more preferable
than $\phi_{m=1}({\bf r})$ at a smaller interaction strength.

\begin{figure}[htp]
\begin{centering}
\includegraphics[clip,width=0.45\textwidth]{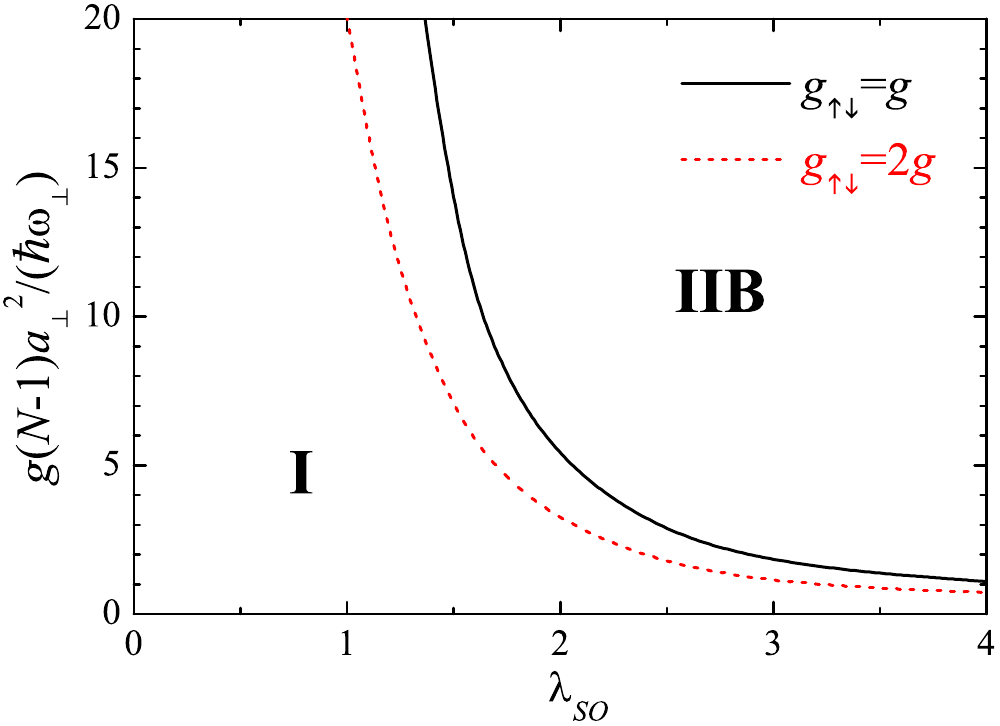} 
\par\end{centering}

\caption{(color online). Phase diagram at $g_{\uparrow\downarrow}=g$ and $g_{\uparrow\downarrow}=2g$.
The critical interaction strength has been shown as a function of
SO coupling.}

\label{fig11} 
\end{figure}

An accurate determination of the threshold could be obtained by monitoring
the instability in a particular collective mode. As the condensate
state may preserve a well-defined parity, we find that the instability
occurs in the lowest quadrupole mode with $m=-2$. In Fig.~\ref{fig10}(b),
we report the lowest quadrupole mode frequency $\omega_{m=-2}$ as
a function of the interaction strength. As the interaction increases,
the real part of mode frequency decreases down to zero and then, the
imaginary part becomes positive, indicating clearly that this mode
will exponentially grow if the condensate is initially in the half-quantum
vortex configuration. The condensate then starts to involve high-order
angular momentum components. The critical interaction strength $g_{c}$
can be simply determined from the softening of the mode frequency,
$\omega_{m=-2}(g=g_{c})=0$.

In Fig.~\ref{fig11}, we present critical interacting strength as
a function of SO coupling at $g_{\uparrow\downarrow}=g$ and $g_{\uparrow\downarrow}=2g$.
The solid line at the isotropic point $g_{\uparrow\downarrow}/g$
has been recently calculated by Xiang-Fa Zhou and Congjun Wu by using
an imaginary time evolution method \cite{WuCPL2011,ZhouPreprint}.
Our results are in excellent agreement with theirs. We find that at
smaller SO coupling the critical interaction strength decreases rapidly
with increasing $g_{\uparrow\downarrow}/g$.


\subsection{Instability against anisotropy in SO coupling strength}

\label{anisoc} 
\begin{figure}[htp]
\begin{centering}
\includegraphics[clip,width=0.45\textwidth]{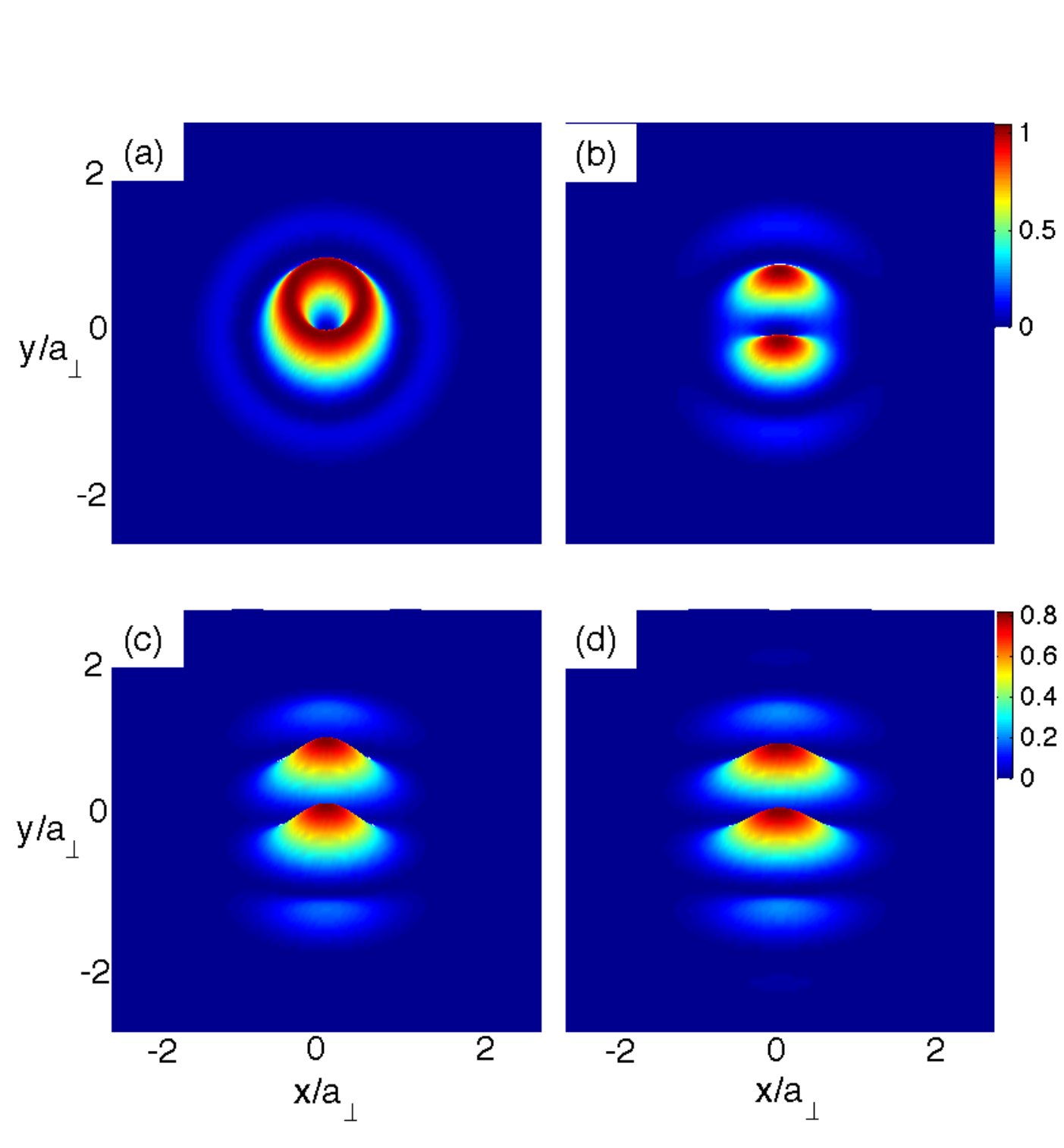} 
\par\end{centering}

\caption{(Color online) Plot of the ground state density profiles of $\downarrow$-spin
component for the parameter set: $g(N-1)=0.1\hbar\omega_{\perp}/a_{\perp}^{2}$,
$g_{\uparrow\downarrow}/g=1.1$, $\lambda_{x}=4.0$, but with varying
ratios of $\lambda_{x}/\lambda_{y}$. (a) Isotropic case: $\lambda_{x}/\lambda_{y}=1.0$,
(b) $\lambda_{x}/\lambda_{y}=1.01$, (c) $\lambda_{x}/\lambda_{y}=1.05$,
(d) $\lambda_{x}/\lambda_{y}=1.1$. Viewing angle is slightly tilted
for aesthetic purposes.}

\label{Gdens_soc} 
\end{figure}

So far we have focused our attention on the half-quantum
vortex state supported by an isotropic 2D harmonic trap subject to
an isotropic Rashba SO coupling. Here we discuss the effect of the
anisotropy in SO coupling strength $\lambda_{R}$ on the stability
of half-quantum vortex state. The effect of the trap anisotropy will
be discussed in the next subsection. In the context of ultracold gases,
anisotropic Rashba spin-orbit coupled was first discussed in Ref.~\cite{GalitskiPRA2008}
and the coupled GP equations were solved for a many-body system in
the absence of the trap and in the restricted scenario when $g_{\uparrow\downarrow}=g$.
Here, we move beyond these restrictions and discuss the ground state
of the system. We write the SO coupling term in the form ${\cal V}_{SO}=-i(\lambda_{y}\hat{\sigma}_{x}\partial_{y}-\lambda_{x}\hat{\sigma}_{y}\partial_{x})$,
where $\lambda_{x},\lambda_{y}$ are SO coupling strengths in the
two perpendicular directions. By including this SO coupling term and
solving the coupled GP equations under the Hamiltonian as given in
Eq.~(\ref{HGP}) using the TSSP technique, we obtain the ground state
wavefunction at various values of anisotropy in SO coupling represented
by $\lambda_{x}/\lambda_{y}$. In Fig.~\ref{Gdens_soc}, we plot
the corresponding ground state density profiles of $\downarrow$-spin
component for an SO coupling strength of $\lambda_{x}=4.0$, and for
various values of $\lambda_{x}/\lambda_{y}$. 

\begin{figure}[h!]
\begin{centering}
\includegraphics[clip,width=0.5\textwidth]{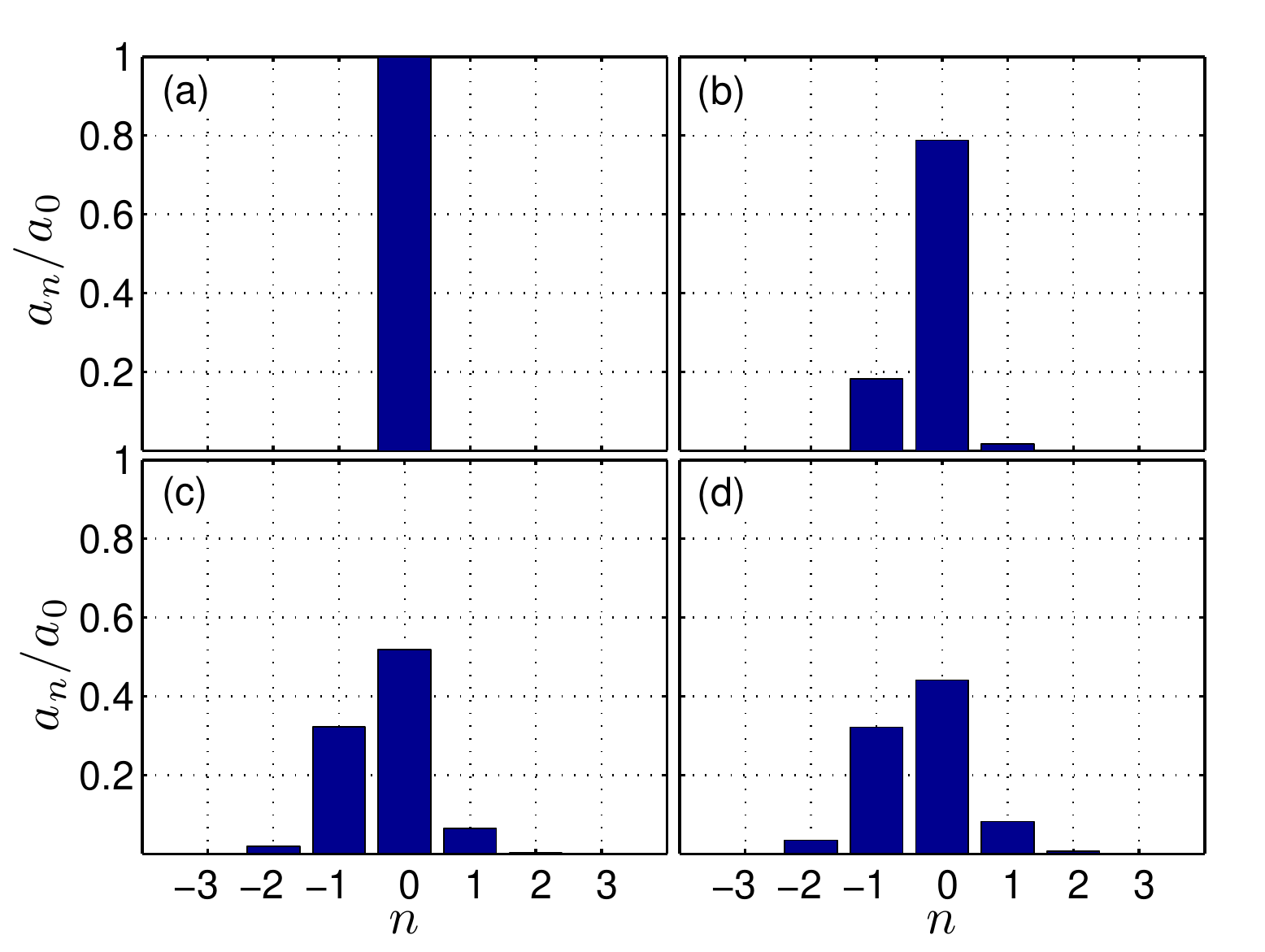} 
\par\end{centering}

\caption{(Color online) Plot of the weights of ground-state wavefunction of
$\downarrow$-spin component - corresponding to the density profiles
in Fig.~\ref{Gdens_soc} - in the $n$th mode. The weights are normalized
with respect to $a_{0}$ computed for half-quantum vortex state with
$\lambda_{x}/\lambda_{y}=1.0$. (a) Isotropic case: $\lambda_{x}/\lambda_{y}=1.0$,
(b) $\lambda_{x}/\lambda_{y}=1.01$, (c) $\lambda_{x}/\lambda_{y}=1.05$,
(d) $\lambda_{x}/\lambda_{y}=1.1$.}

\label{Gdens_soc_bar} 
\end{figure}

We see from Fig.~\ref{Gdens_soc}(a) that the half-quantum vortex
state is indeed the ground state (already mentioned in Fig.~\ref{fig1}(b))
for the parameter set: $g(N-1)=0.1\hbar\omega_{\perp}/a_{\perp}^{2}$,
$g_{\uparrow\downarrow}/g=1.1$, $\lambda_{x}=4.0$ and $\lambda_{x}/\lambda_{y}=1.0$.
We shall now analyze the pattern in which the density profile changes
with anisotropy in SO coupling strength as shown in Fig.~\ref{Gdens_soc}(b)-(d).
It is evident from the density distributions in Fig.~\ref{Gdens_soc},
that the half-quantum vortex state is unstable even against small
anisotropy in SO coupling strength. Adopting a similar method as presented
in Ref.~\cite{dipole}, we analyze this systematically by expanding
the wavefunction of $\downarrow$-component in an orthogonal basis
set of the form: $\Phi_{\downarrow}(\rho)=\Sigma_{n}\, f_{n}(\rho)\, e^{i\,(2n+1)\,\varphi}$,
where $n$ measures the vorticity, and $f_{n}(\rho)$ absorbs the
$n$th mode's contribution in radial direction. We quantify the weights
of the wavefunction in the $n$th mode by computing $a_{n}=\int d\rho\,\vert f_{n}(\rho)\vert^{2}$.
In Fig.~\ref{Gdens_soc_bar}, we plot the weights $a_{n}$ relative
to $a_{0}$ computed for half-quantum vortex state with $\lambda_{x}/\lambda_{y}=1.0$.
As we would expect, for this isotropic case, $a_{0}=1$ and $a_{n}=0$
for $n\neq0$. As anisotropy in SO coupling strength increases, more
and more $n\neq0$ components will be mixed into the ground state.


\subsection{Instability to anisotropy in trap potential}

\label{anitrap} 
\begin{figure}[htp]
\begin{centering}
\includegraphics[clip,width=0.45\textwidth]{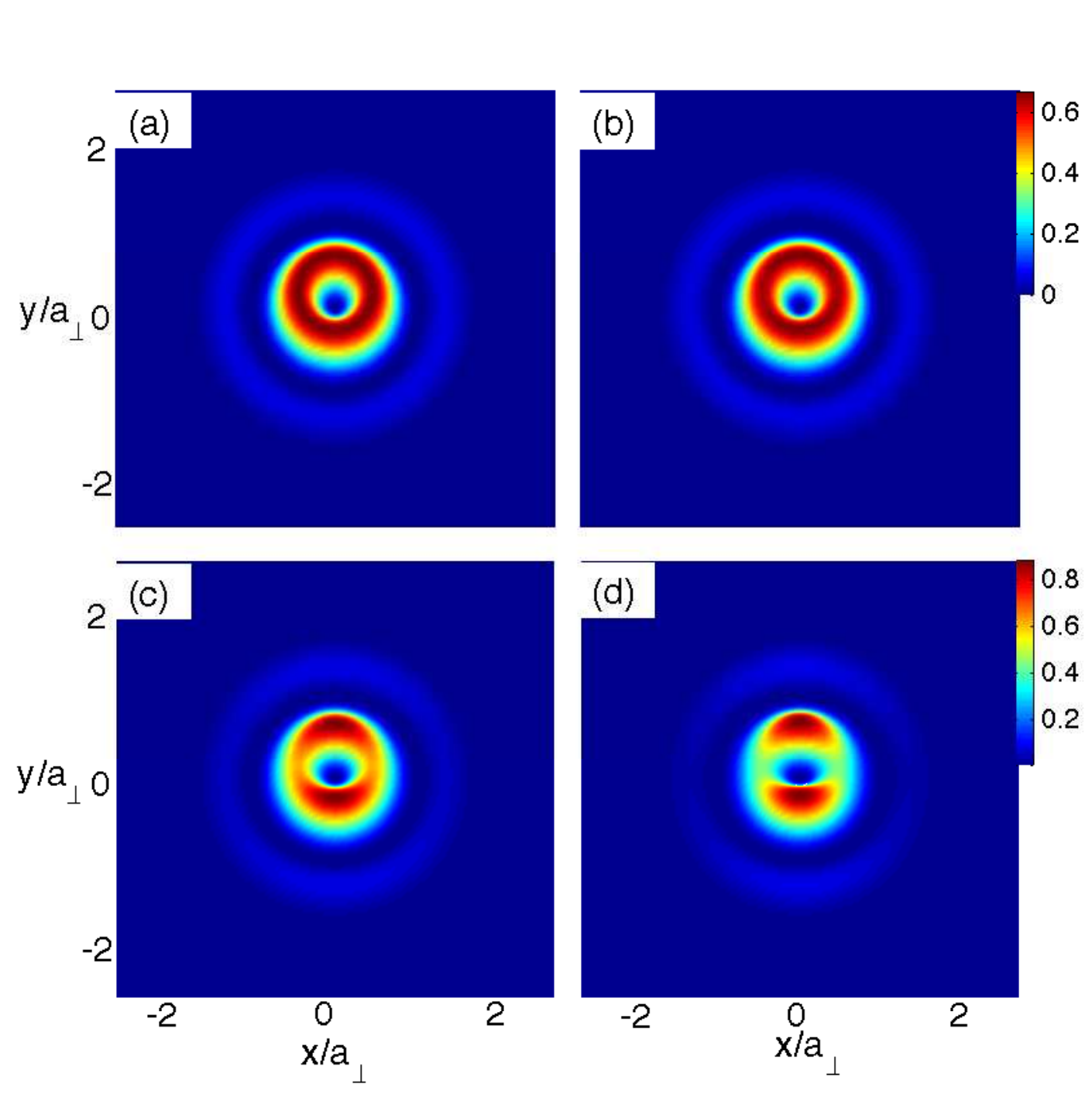} 
\par\end{centering}

\caption{(Color online) Plot of the ground state density profiles of $\downarrow$-spin
component for the parameter set: $\lambda_{SO}=4.0$, $g(N-1)=0.1\hbar\omega_{\perp}/a_{\perp}^{2}$,
$g_{\uparrow\downarrow}/g=1.1$, but with varying ratios of $f_{y}=\omega_{y}/\omega_{x}$.
(a) Isotropic case: $f_{y}=1.0$, (b) $f_{y}=1.01$, (c) $f_{y}=1.05$,
(d) $f_{y}=1.1$. Viewing angle is slightly tilted for aesthetic purposes.}

\label{Gdens_trap} 
\end{figure}

Now we examine the effect of anisotropy in the trapping
potential, but with isotropic SO coupling, on the stability of half-quantum
vortex state. We write the trapping potential in the form $V(x,y)=M(\omega_{x}^{2}x^{2}+\omega_{y}^{2}y^{2})/2=M\omega_{\perp}^{2}(x^{2}+f_{y}^{2}y^{2})/2$,
where $\omega_{x}=\omega_{\perp},\omega_{y}=f_{y}\omega_{\perp}$
are trapping frequencies in $x$- and $y$-directions respectively.
We again obtain the ground state wavefunctions at various values of
$f_{y}$ by solving the coupled GP equations using the TSSP technique.
In Fig.~\ref{Gdens_trap}, we plot the corresponding ground state
density profiles of $\downarrow$-spin component for an SO coupling
strength of $\lambda_{SO}=4.0$, and for various values of trap anisotropy
ranging from 0 to 10$\%$. 

\begin{figure}[h!]
\begin{centering}
\includegraphics[clip,width=0.5\textwidth]{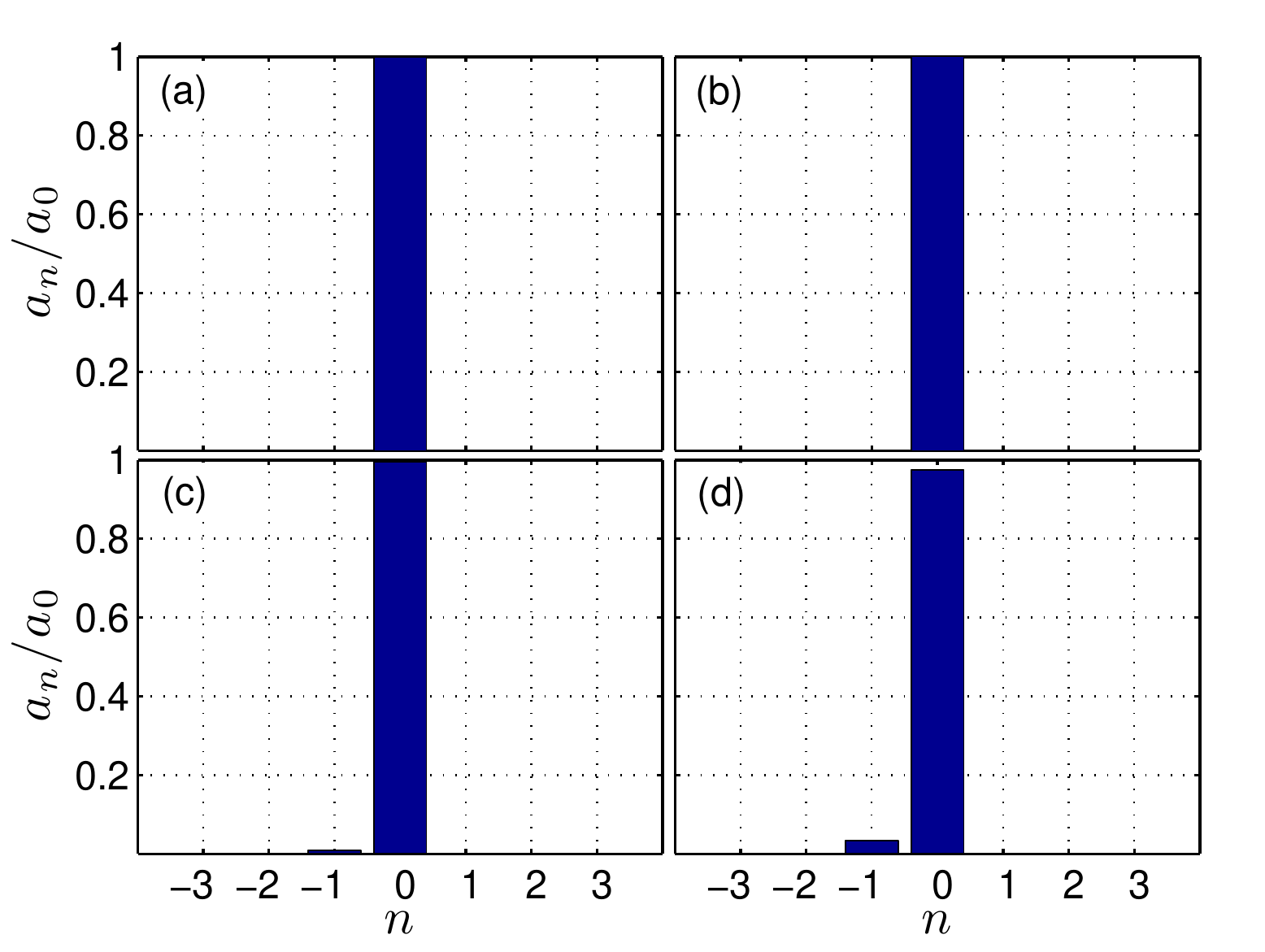} 
\par\end{centering}

\caption{(Color online) Plot of the weights of ground-state wavefunction of
$\downarrow$-spin component - corresponding to the density profiles
in Fig.~\ref{Gdens_trap} - in the $n$th mode. The weights are normalized
with respect to $a_{0}$ computed for half-quantum vortex state with
$f_{y}=1.0$. (a) Isotropic case: $f_{y}=1.0$, (b) $f_{y}=1.01$,
(c) $f_{y}=1.05$, (d) $f_{y}=1.1$.}

\label{Gdens_trap_bar} 
\end{figure}

We see from Fig.~\ref{Gdens_trap}(a) that the half-quantum vortex
state is indeed the ground state (already mentioned in Fig.~\ref{Gdens_soc}(a))
for the parameter set: $\lambda_{SO}=4.0$, $g(N-1)=0.1\hbar\omega_{\perp}/a_{\perp}^{2}$,
$g_{\uparrow\downarrow}/g=1.1$. We shall now analyze the pattern
in which the density profile changes with trap anisotropy Fig.~\ref{Gdens_trap}(b)-(d).
It is evident from the density distributions in Fig.~\ref{Gdens_trap},
that the vortex core becomes increasingly anisotropic with increasing
$f_{y}$. We analyze this systematically by expanding the wavefunction
of $\downarrow$-component in an orthogonal basis set and quantifying
the weights in the $n$th mode by $a_{n}$, as mentioned in Sec.~\ref{anisoc}.
In Fig.~\ref{Gdens_trap_bar}, we plot the weights $a_{n}$ relative
to $a_{0}$ computed for half-quantum vortex state with $f_{y}=1.0$.
As we would expect, for the isotropic case with $f_{y}=1.0$, $a_{0}=1$
and $a_{n}=0$ for $n\neq0$. As trap anisotropy increases, we observe
that the ground state is a mixture of $n\neq0$ components as well.
Nevertheless, we see that the trap anisotropy has a much smaller effect
on the half-quantum vortex state than the anisotropy in the SO coupling
strength. 


\section{Conclusions}

In summary, we have investigated systematically the ground condensate
state of a spin-orbit coupled spin-1/2 Bose gas confined in two-dimensional
harmonic traps. The density distributions and collective density excitations
have been obtained respectively by solving the Gross-Pitaevskii equation
and Bogoliubov equation, which are generalized to include the spin-orbit
coupling. We have found that:

(1) The condensate is in a half-quantum vortex state, if the intra-species
interaction $g$ is smaller than inter-species interaction $g_{\uparrow\downarrow}$
and, if the interaction strength is below a threshold $g_{c}$. We
have calculated the threshold by monitoring the unstable quadrupole
mode with an azimuthal angular momentum $m=-2$. A phase diagram for
the half-quantum vortex state is therefore determined, as given in
Figs.~\ref{fig1} and \ref{fig11}.

(2)The half-quantum vortex state (the phase I) will turn into a superposition
of two degenerate half-quantum vortex states (the phase IIA) if $g>g_{\uparrow\downarrow}$
and will start to involve high-order angular momentum components (the
phase IIB) if $g>g_{c}$, where $g_{c}$ depends critically on the
ratio $g_{\uparrow\downarrow}/g$. The half-quantum vortex state is
unstable against small anisotropy in SO coupling strength and large
anisotropy in trapping potential. The state tends to be a superposition
of higher angular momentum states.

(3) In the presence of spin-orbit coupling, the behavior of collective
density modes becomes complicated. In particular, the breathing mode
with $\omega=2\omega_{\perp}$ and the dipole mode with $\omega=\omega_{\perp}$
are no longer the exact solutions of the many-body system.

(4) The condensate wave-functions in the phases IIA and IIB are yet
to be determined using the time-splitting spectral method for GPE.
These wave-functions break the rotational symmetry. We anticipate
that interesting density patterns will emerge in the limit of very
large interatomic interactions. This is to be explored in future studies.
\begin{acknowledgments}
We would like to thank Congjun Wu and Xiang-Fa Zhou for useful discussions.
BR thanks Lin Dong and Hong Lu for useful discussions. HH and XJL
was supported by the ARC Discovery Projects No. DP0984522 and No.
DP0984637. HP is supported by the NSF, the Welch Foundation (Grant
No. C-1669) and the DARPA OLE program. \end{acknowledgments}

\end{document}